\documentclass[11pt]{article}
\usepackage{graphicx} 

\usepackage{url}
\usepackage{authblk} 

\title{On Signatures of a Possible New Physics Resonance in Atmospheric Air Showers Using a Parameterized Model}

\date{\today}

\author[a]{Ji\v{r}\'{\i} Kvita\thanks{Corresponding author: \texttt{jiri.kvita@upol.cz}}}

\affil[a]{Palacky University Olomouc, Faculty of Science, Joint Laboratory of Optics of Palacky University and Institute of Physics of the Czech Academy of Sciences, 17. listopadu 1192/12, 779 00 Olomouc, Czech Republic}

\usepackage{xcolor}

\usepackage[utf8]{inputenc}
\usepackage{slashed}
\usepackage{amsmath}
\usepackage{graphicx,epsfig}
\providecommand{\Conex}{\textsl{Conex}}
\providecommand{\logE}{\ensuremath{\log_{10} E / \mathrm{eV}}}
\providecommand{\Xmax}{\ensuremath{X_\mathrm{max}}}
\providecommand{\sigmaX}{\ensuremath{\sigma_{X_\mathrm{max}}}}
\providecommand{\gcm}{\ensuremath{\mathrm{g/cm}^2}}

\begin{document}

\maketitle

\begin{abstract}
    We present a parameterized model of atmospheric particle showers initiated by cosmic rays. Few physics shower parameters are tuned in a comparison to the Conex generator. Resulting shower properties are studied, with a comment on the cases where multiple shower maxima develop.
    Finally, we implement simple models of new physics resonance of masses of 100 GeV and 1 TeV and examine their effects on the shower profile, depth and maximum variation in dependence of the decay channel of the hypothetical resonance. It is shown that a new resonance effects can appear at the energy threshold and can persist for about a decade in $\log_{10} E/\mathrm{eV}$.
    Various assumed decay modes of the hypothetical resonance have different effects on the direction and shape of the modified average shower depth as function of the energy, with possible implications for current or future measurements. It is shown that, within the presented model, the visibility of the resonance in modified shower depth strongly depends on the resonance width. A significant modification at 10\% width gradually diminishes towards the percent-level width. We propose that looking at the 2D distributions of the two first individual shower moments can also reveal signatures of new physics.
\end{abstract}


\section{Introduction}

Cosmic rays observed as extensive air showers (EAS) of particles developing in the Earth's atmosphere provide the message of the very existence of these cosmic rays. Their observation raises the intriguing questions on their origin, acceleration mechanism, effects shaping their spectra, and on the composition of the primary cosmic rays.
With their energies exceeding by far those accessible by current man-made particle accelerators, the question arises whether possible signs of new physics could be observed in the measured showers properties.
A long-standing question of the muon content of the observed showers not being described by current models is one of the examples~\cite{PierreAuger:2014ucz,Albrecht:2021cxw}.
Still, the hadronization and interaction models applied to the showers of the cosmic rays of ultra-high energies are based on the extrapolation of the observed processes at lower energies at particle accelerators, with mechanisms like rescattering~\cite{Pierog:2025ixr} being proposed to solve some of the observed puzzles.
The accelerator data are often based on non-ideal particle species colliding, \emph{e.g.} proton-proton, lead-lead or proton-lead collisions, with the long-awaited proton-oxygen run of the LHC taken only in 2025.

Here we explore the possibility of a a hypothetical particle resonance of various masses and widths. While some new physics effects in the observables related to the air showers have already been explored \emph{e.g.} in~\cite{Brooijmans:2016lfv} or~\cite{Fischer:2020dmn}, we focus on examining the threshold effects and study the resonance persistence for various widths and decay modes as function of the primary particle energy.

\section{The parameterized simulation}

The processes within the proton or nucleus-initiated shower development are of two distinct kinds: the truly hadronic sub-shower and the electromagnetic (EM) one, the latter an inevitable integral part of the full shower even if initiated by a primary hadron, mostly due to neutral mesons like $\pi^0$ or $\eta$ produced within the shower and decaying electromagnetically.
There are of course other more subtle processes, ranging from neutron interactions to photonuclear or muon-nuclear interactions, which are sometimes being neglected, depending on a model.

The challenge in simplified models arises in how to reproduce the known shower maximum profiles using particles splitting schemes, where each particle energy is split usually evenly to the process-dependent products each radiation or interaction length, and at the same time retain some realistic fluctuations in the shower development.

We extend the traditional simplified approach based on splitting each particle in the shower every characteristic length into two particles sharing evenly the energy of the parent particle, motivated by allowing more natural fluctuations in the shower development and energy share within the shower particles. Compared to~\cite{Ulrich:2010rg} we do not attempt to improve the hadronic sub-shower description but we aim to use a reasonably realistic model to study effects of possible new physics.

\begin{figure}[!ht]
    \centering
    \includegraphics[width=0.6\linewidth]{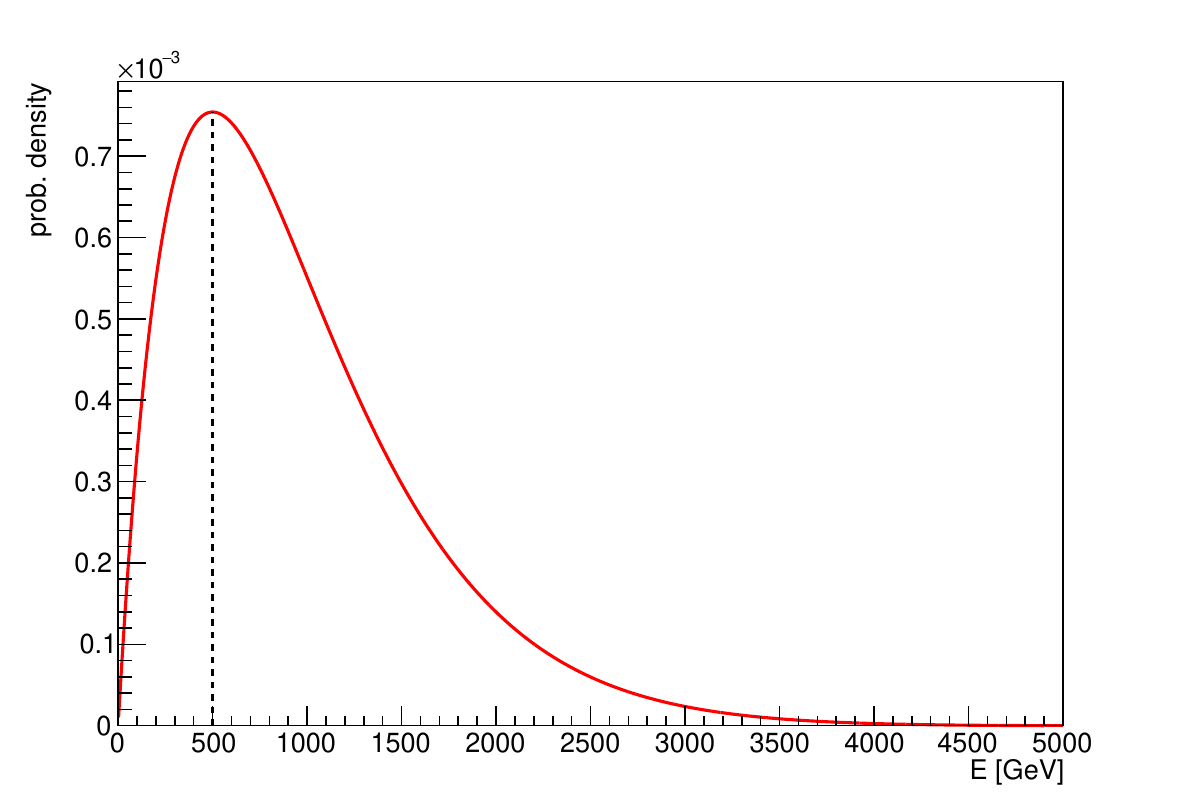}
    \caption{The probability density function used to share energies of the parent particle between the produced hadrons. The vertical line indicates the average energy per one pion, here on the example of $N=10$ and the parent hadron energy fraction to be shared by the charged pions of 5~TeV.}
    \label{fig:pdf_Epions}
\end{figure}

\begin{figure}[p]
    \centering
    \includegraphics[width=0.8\linewidth]{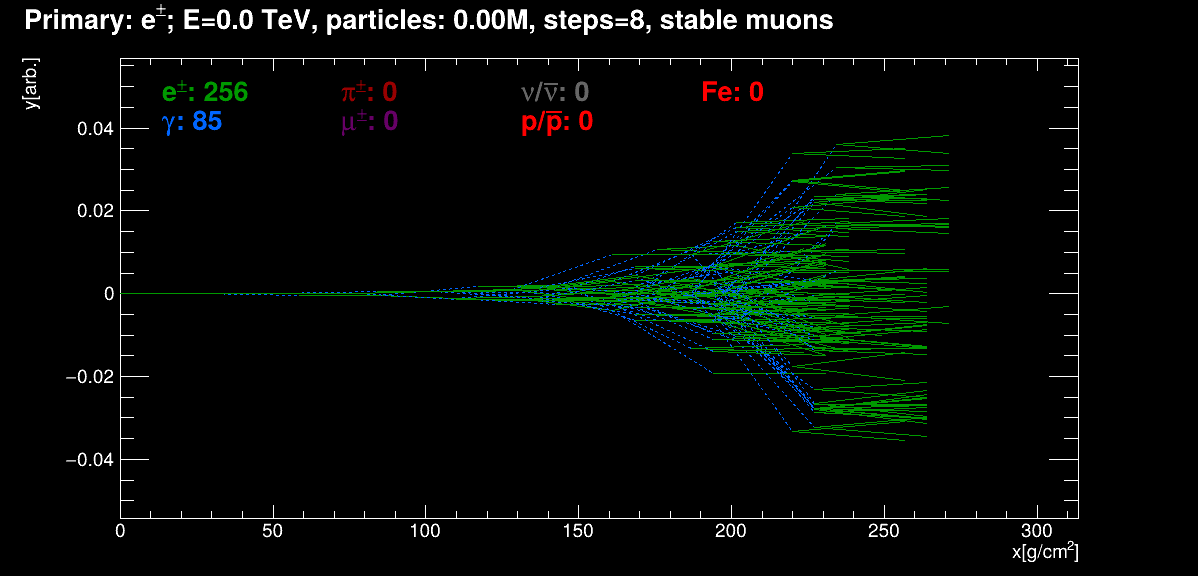} \\ 
    \includegraphics[width=0.8\linewidth]{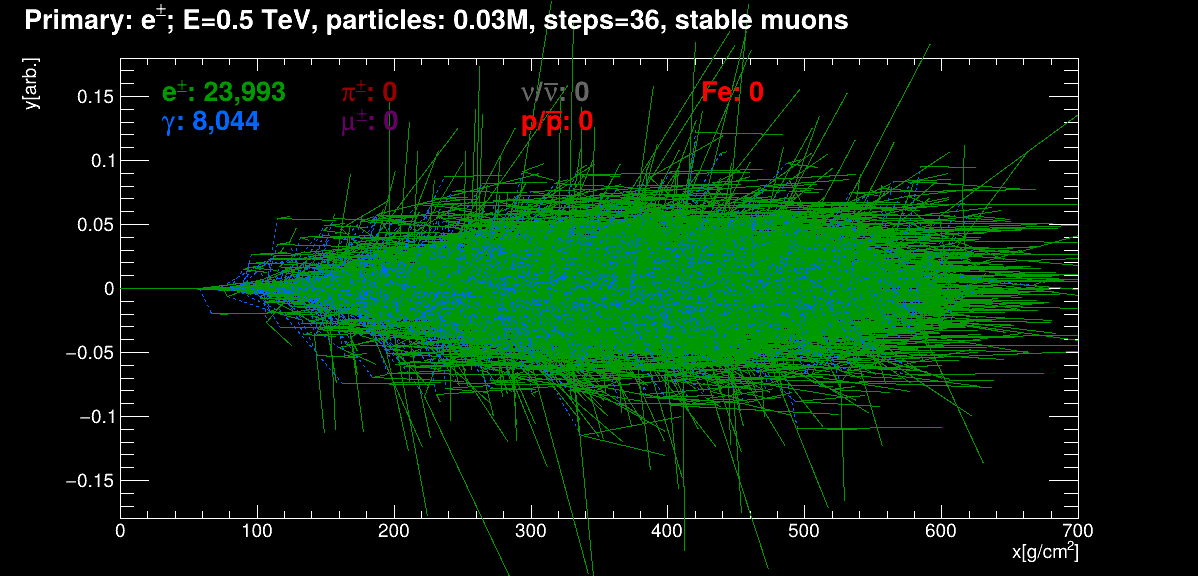} \\ 
    \caption{Two visualizations of a simple parameterized atmospheric air shower, once splitting particles every half-interaction or radiation lengths (top), or following a random choice from the exponential probability distributions (bottom). The showers are initiated by a 100~GeV and 500~GeV electrons, respectively. Total particles multiplicities are indicated in the legend. Shower transverse profile is only indicative, see text. Colors in legend correspond to colors of particle tracks in the shower.}
    \label{fig:airsim_steps}
\end{figure}

\begin{figure}[p]
    \centering
    \includegraphics[width=0.8\linewidth]{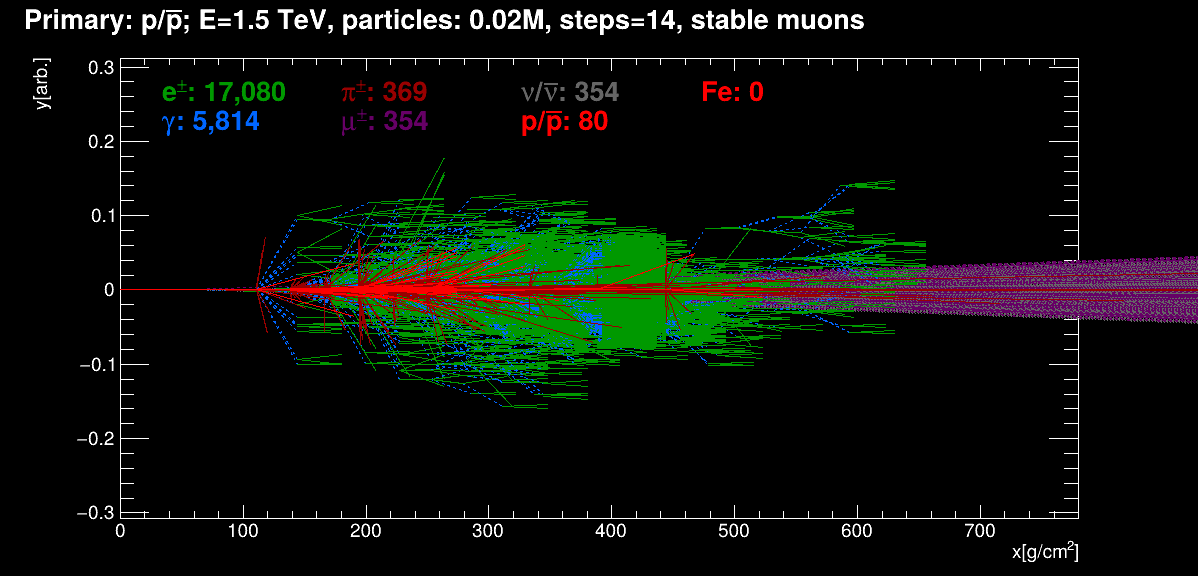} \\ 
    \includegraphics[width=0.8\linewidth]{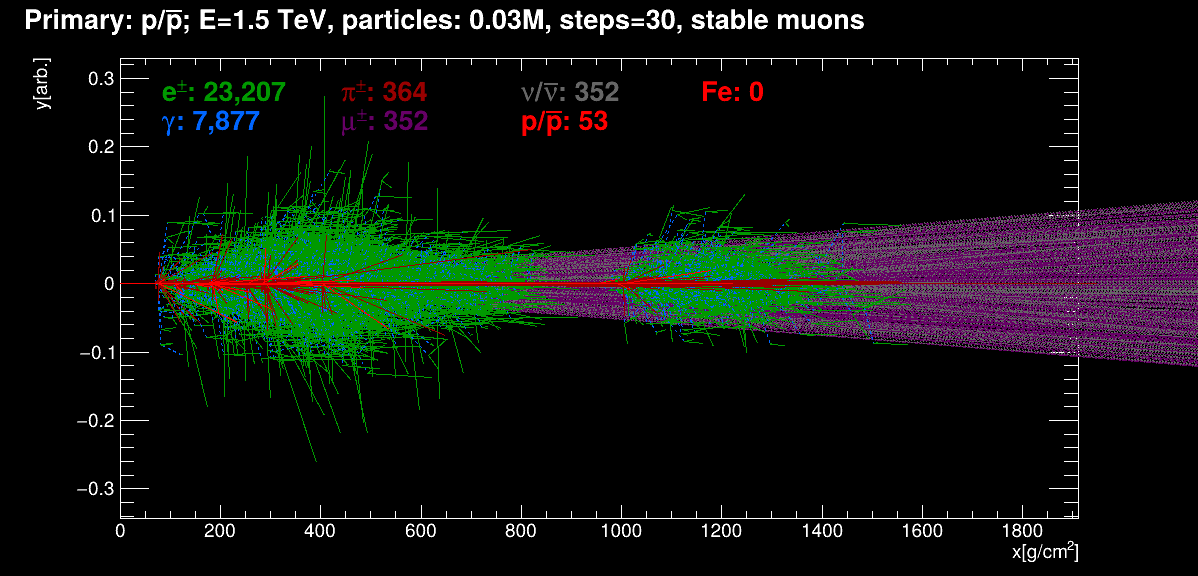} \\ 
    \includegraphics[width=0.8\linewidth]{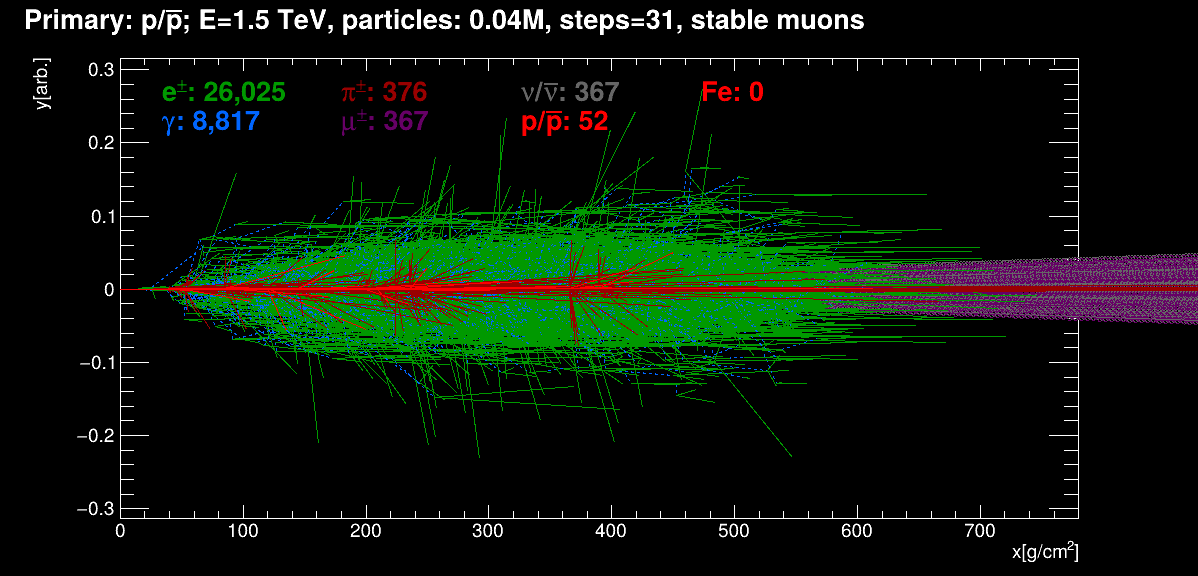} \\ 
    \caption{Three visualizations of a simple parameterized atmospheric air shower, once splitting particles every half-interaction or radiation lengths (top), or following a random choice from the exponential probability distributions (middle and bottom). The showers are initiated by a 1.5~TeV protons with the middle shower exhibiting a double peak structure. Total particles multiplicities are indicated in the legend and differ for each splitting model. Shower transverse profile is only indicative, see text. Colors in legend correspond to colors of particle tracks in the shower.}
    \label{fig:airsim_steps_p}
\end{figure}

We thus implement an intermediate model of air showers initiated by a proton, iron or an electron or photon developed and implemented in Python3~\cite{githubjk}. We consider the following processes:
\begin{itemize}
    \item production of a leading proton in hadronic interactions, governed by the hadronic (proton or pion) interaction length and the inelasticity parameter~\cite{MATTHEWS2005387} being the energy fraction taken by the leading proton; the rest of the energy used to produce pions or additional protons;
    \item production of a number of both charged and neutral pions, neutral ones decaying immediately into two photons;
    \item radiation losses (bremsstrahlung) in terms of photons radiated from electrons and positrons based on the radiation length in air.
    \item photons conversion to $e^+ e^-$ pairs based on the  $\frac{9}{7}$ of the radiation length in air.
    \item We neglect ionization losses, neutrons production and their interactions as well as photonuclear or muon nuclear interactions etc. We also neglect diffractive processes.
\end{itemize}
All processes are governed by exponential ``decay'' law formulas with characteristic lengths. This way, the meaning of the lengths as mean travel or interaction paths is kept while also allowing processes with earlier or later interactions, namely longer travel paths, leading also to cases where multiple shower maxima can develop. A comparison for electromagnetic showers is show using regular splitting each interaction half-lengths or the exponential form (see~Figure~\ref{fig:airsim_steps}).
Similar is show for proton-initiated showers in~Figure~\ref{fig:airsim_steps_p}. Total particle counts in showers fluctuate but also differ due to the different splitting models.
The following two splitting schemes naturally arise:
\begin{itemize}
    \item Split the electromagnetic particle (photon, electron or positron) every relevant half-length and split the energy evenly to  the particles produced; or
    \item Allow shower longitudinal variations by choosing an interaction point according to the exponential distribution governed by the characteristic length of the process. In this case, split the particle into two with energy fractions $x$ and $1-x$ with $x$ being distributed according to
    \begin{enumerate}
        \item the fractions as predicted by the cross section of the process;
        \item the exponential law governed by the characteristic length of the process.
        \item a mixed scenario by choosing the interaction point from the exponential distribution and the energy fraction from the corresponding cross-section formula.
    \end{enumerate}
    It turns out that in all these cases, fluctuations in the random choices from the exponential lead to longitudinal extensions of the electromagnetic showers which are not compatible with other models. We thus optimizes the maximal interaction point to be a multiple ($k$) of the typical process length, arriving to 
    \begin{enumerate}
        \item $k = 1.25$ for uniform energy splitting;
        \item $k = 1.125$ for the case when using the exponential law.
    \end{enumerate}
    We eventually use the mixed approach, \emph{i.e.} we allow fluctuations in the depth development of the EM shower, but only to the extend so that the shower depth is compatible with reality, and use the process-specific fraction distribution of the parent particle energy. Namely, we split the photon into an electron-positron pair with fractions according the $\sim \frac{9}{7} (1 - \frac{4}{3}  x (1  - x) ) $ distribution and the resulting electron energy fraction after radiating a photon distributed as $\sim (\frac{4}{3} - \frac{4}{3} (1-x) + (1-x)^2)$, the well-known leading-order kinematics of the related cross-sections.
    \item Similar maximal interaction point is designed also for hadronic processes, where, in the end, we do not impose any restriction, see below the section dedicated to the parameters tuning.

    \item Hadrons produced in the interaction of an initial hadron of energy $E_0$ are considered as pions, allowing a substitution of several charged pions by additional protons according to the Poisson distribution with mean of 1. The energy $E$ of each produced charged or neutral hadron is drawn from a distribution of the form (see also Figure~\ref{fig:pdf_Epions})
    $$f(E|E_0,\lambda_E) = \mathcal{A} \, E (E_0 - E) \, e^{-E/\lambda_E}\,, \quad E \in (0, E_0) \,,$$
    where the $\lambda_E$ parameter is set so that the mode of the distribution equals $E / N$ where $N$ is the number of produced hadrons. 
    It turns out that
   \begin{align}
        E_\mathrm{mode} &\equiv E/N = \lambda_E + \frac{E_0}{2} - \sqrt{\lambda_E^2 + \left( \frac{E_0}{2}\right)^2} \\
     \lambda_E &= E\,\frac{N-1}{N(N-2)} \\ 
     \mathcal{A}^{-1} &= 2\lambda_E^3 \left(1 - e^{-E_0/\lambda_E} \right) - \lambda_E^2 E_0 \left(1 + e^{-E_0/\lambda_E} \right)\,.  
    \end{align}
    We allow for a maximum of 100 iterations to generate hadrons according to this law to gradually add up their energies to the one of the initial hadron, bailing out at this condition and assigning the remaining energy to the last hadron.

\end{itemize}

The number of particles present in the shower is recorded as function of the atmospheric depth $x$ in $\mathrm{g/cm}{}^2$ from the decay point of the primary particle. This way, the ambiguity stemming from the fluctuations in the start of the shower development is removed.
Motivated by the experimental detection techniques based on the fluorescence light, we only count particles with substantial ionization losses, \emph{i.e.} only protons, charged pions; and electrons or positrons.

\begin{figure}[p]
    \centering
    \includegraphics[width=0.7\linewidth]{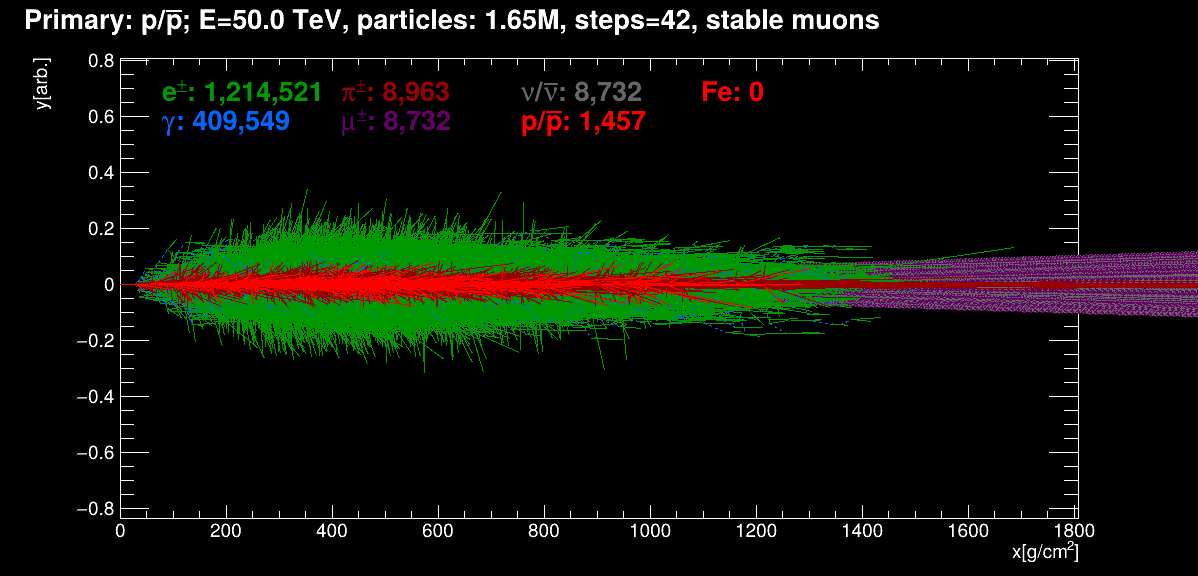}  
    \includegraphics[width=0.29\linewidth]{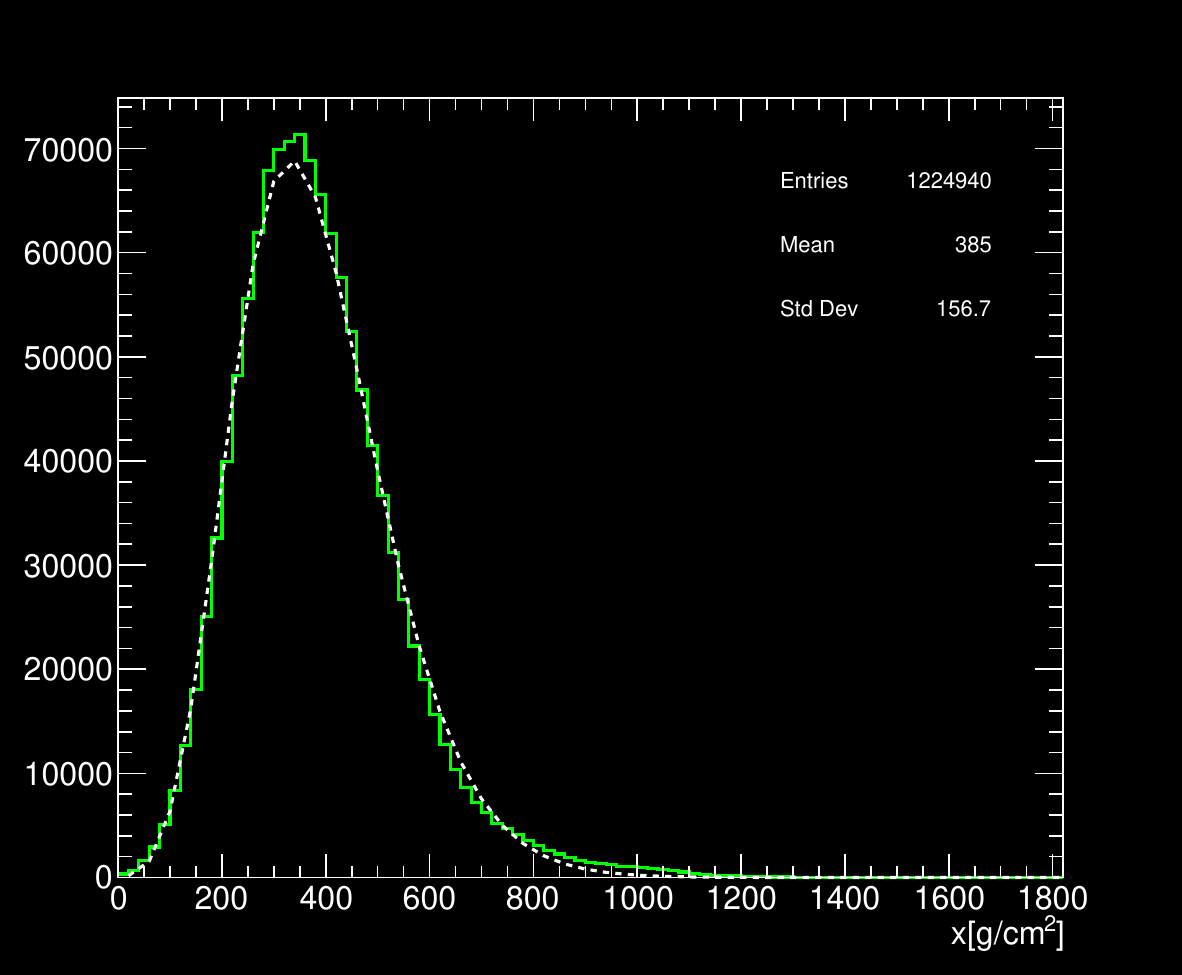} \\
    \includegraphics[width=0.7\linewidth]{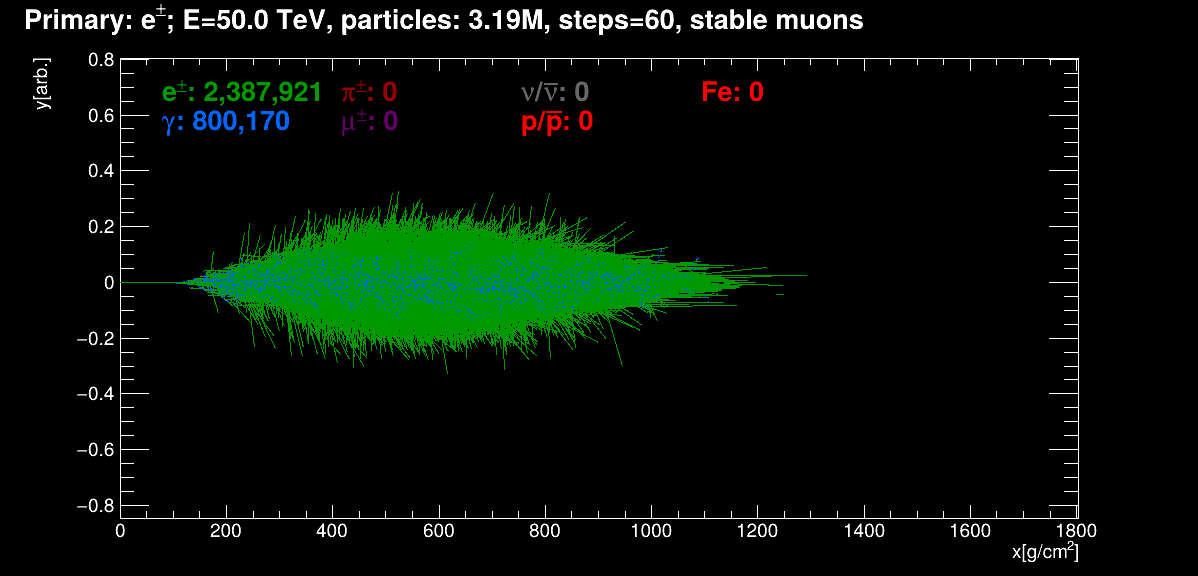}  
    \includegraphics[width=0.29\linewidth]{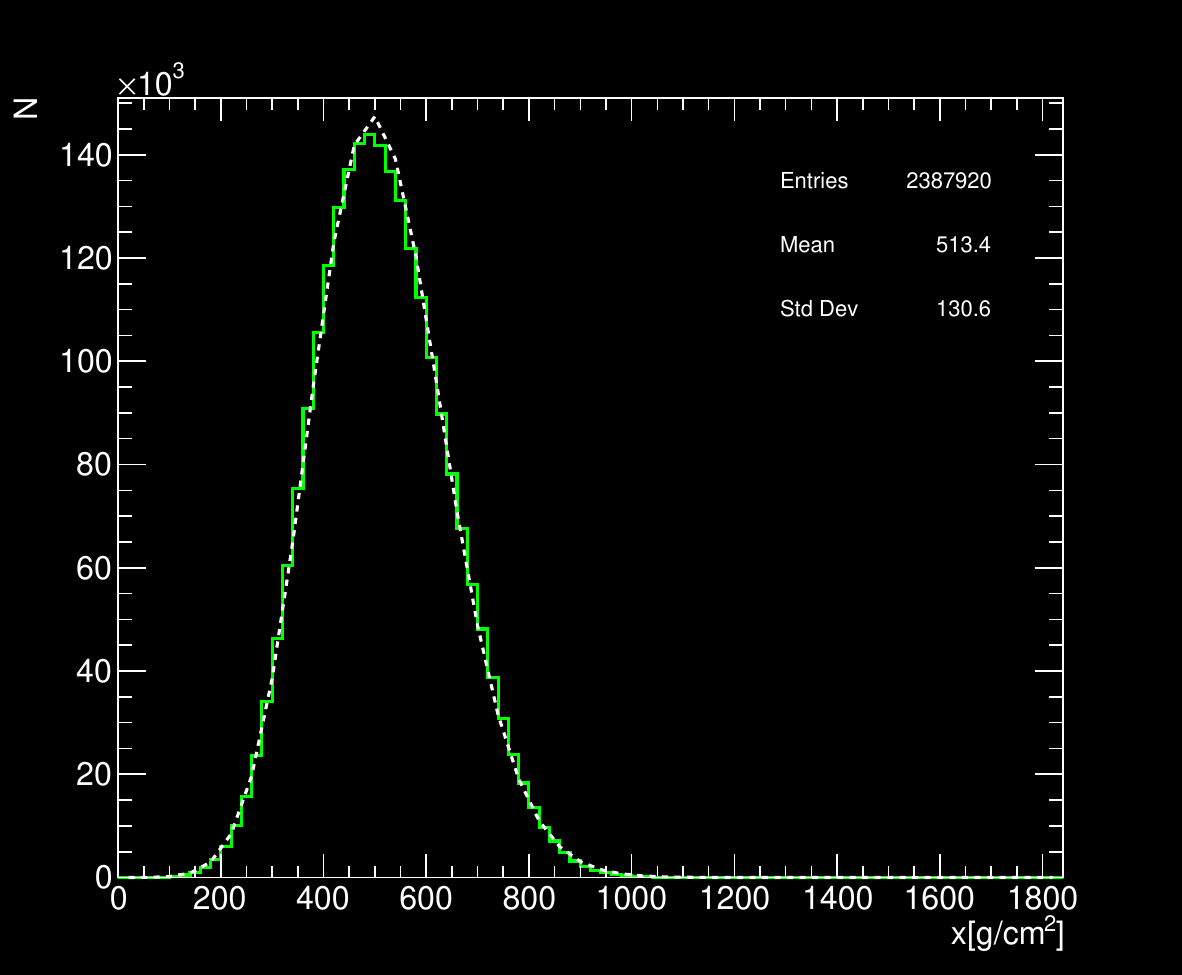}
    \caption{Top left: A visualization of a simple parameterized atmospheric air shower initiated by a 50~TeV proton, with stable muons assumed, and particles multiplicities indicated in legend. Only the EM and hadronic component is propagated, with neutrinos from pion decays also shown. Neutrons are not taken into account. Shower transverse profile is only indicative, see text. Top right: Longitudinal shower development in terms of the number of charged particles as function of the atmospheric depth. A fit based on the Gaisser-Hillas profile is shown. Similar is shown in the bottom row for an electromagnetic shower initiated by an electron of the same initial energy.}
    \label{fig:airsim_example}
\end{figure}

Longitudinal shower profiles are usually fitted by the ansatz function by Gaisser and Hillas~\cite{gaisser1977reliability}
$$ \frac{\mathrm{d}N}{\mathrm{d}x} = N_0 \left( \frac{x-x_0}{X_\mathrm{max}-x_0}\right)^\frac{X_\mathrm{max}-x_0}{\lambda} \, \exp{\left(-\frac{x-X_\mathrm{max}}{\lambda}\right)}$$
for $x > x_0$. 
The fit formula (see Figure~\ref{fig:airsim_example} for an example) describes the simulated events very well, which is indicative of the simple parameterized simulation describing the essentials of the shower longitudinal development.

However, in order to avoid fit instabilities over thousands of showers, 
we perform a simple Gaussian fit around the maximal bin and take such fitted Gaussian mean as the measure of the maximum of the shower development, \Xmax{}.
In case of the \Conex{} showers we simply take the \Xmax{} as already provided by the generator for each shower, having checked that the Pearson correlation factor between the fitted peak position and the \Xmax{} as provided by \Conex{} is above 96\%.

Figure~\ref{fig:airsim_example} compares two showers initiated by a proton or an electron of the same primary energy, revealing the important fact that the electromagnetic showers develop slightly later but more importantly more deeply into the medium, leading to a later maximum of the shower development.

\section{Shower parameters tuning}

Following the parameters introduced in \cite{MATTHEWS2005387} we also  introduce the per-collision parameters governing the multiplicity of produced particles but we extend the model to allow these parameters fluctuate event-by-event. We do not use the freedom of allowing them to vary also with energy. Although this would be an easy extension, it was not necessary for the study.
These shower development parameters also allow for fluctuations in the longitudinal shower development and affect the standard deviation of the shower maximum, \sigmaX{}, evaluated over the showers.
The following private parameters have been used in tuning and varied as indicated.

\begin{itemize}
    \item Inelasticity
    \begin{itemize}
       \item Meaning: energy fraction of the colliding hadron going to the production of the leading proton, the rest shared by charged and neutral pions and possibly also additional non-leading protons.
        \item Initial value: a random number drawn from the Gaussian distribution centered at 0.45 (for tuning varied between 0.4 and 0.5 in steps of 0.05), with standard deviation of 0.2.
    \end{itemize}
    \item Number of charged hadrons $C$ produced in an interaction of a proton or a pion:
    \begin{itemize}
        \item 
        $C$ chosen from the Gaussian distribution of mean of 10 (varied from 2 to 12 in steps of 2) and width of 3 (varied also to 2 and 4 in the tuning procedure).
    \end{itemize}
    \item The fraction of neutral pions produced: randomly chosen from uniform distribution between 1/4 and 1/3; not systematically varied. 

\end{itemize}

The material and physics parameters of the atmosphere and the relevant processes were taken as follows:
\begin{itemize}
    \item Critical energy below which the hadrons production is terminated, $E_C = 20\,$GeV;
    \item Critical energy for electrons below which bremsstrahlung is terminated, $87.92\,$MeV;
    \item Radiation length $X_0 = 37 \,\gcm{}$, in air;
    \item Photons interaction length $\lambda_\gamma = \frac{9}{7} X_0$;
    \item Inelastic pion interaction length $\lambda_\pi = 120 \,\gcm{}$;
    \item Inelastic hadronic interaction length $\lambda_I = 80  \,\gcm{}$.
\end{itemize}
For an overview, the corresponding cross-section for the processes used are listed in Table~\ref{tab:xsects}, computed from the radiation and interaction lengths (in $\mathrm{g/cm}^2$) as 
$$ \sigma = \frac{1}{\rho\lambda} \frac{A}{N_A}$$
with dry air density $\rho = 1.233 \times 10^{-3}\,\mathrm{g/cm}^3$ and $A = 28.96$.

\begin{table}[!h]
    \centering
    \begin{tabular}{l|r|r|r}
Process in air & length [$\mathrm{g/cm}^2$]& length [m] & cross-section [mb] \\ \hline
Electron radiative losses           & 37.0 $\mathrm{g/cm}^2$ &  300 m  &   1300 mb \\
Photon conversion                   & 47.6 $\mathrm{g/cm}^2$ &  386 m  &   1011 mb \\
Pion nuclear interaction length             & 120.0 $\mathrm{g/cm}^2$ &  973 m  &   401 mb \\
Proton nuclear interaction length           & 80.0 $\mathrm{g/cm}^2$ &  649 m  &   601 mb \\
    \end{tabular}
    \caption{Various processes and their interaction lengths and corresponding cross-sections in air at standard conditions.}
    \label{tab:xsects}
\end{table}

\begin{figure}
    \centering
    \includegraphics[width=0.72\linewidth]{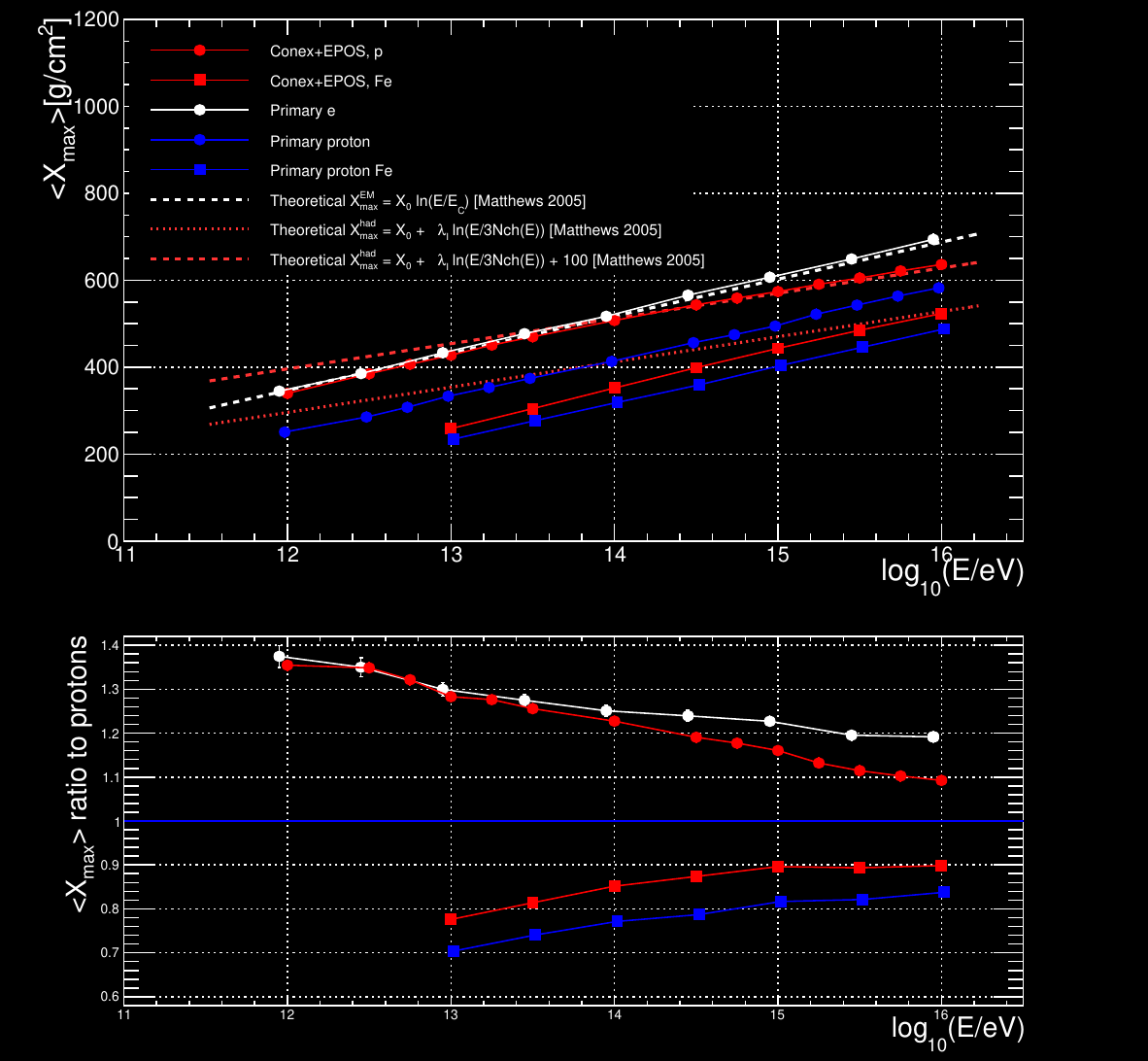}
    \includegraphics[width=0.72\linewidth]{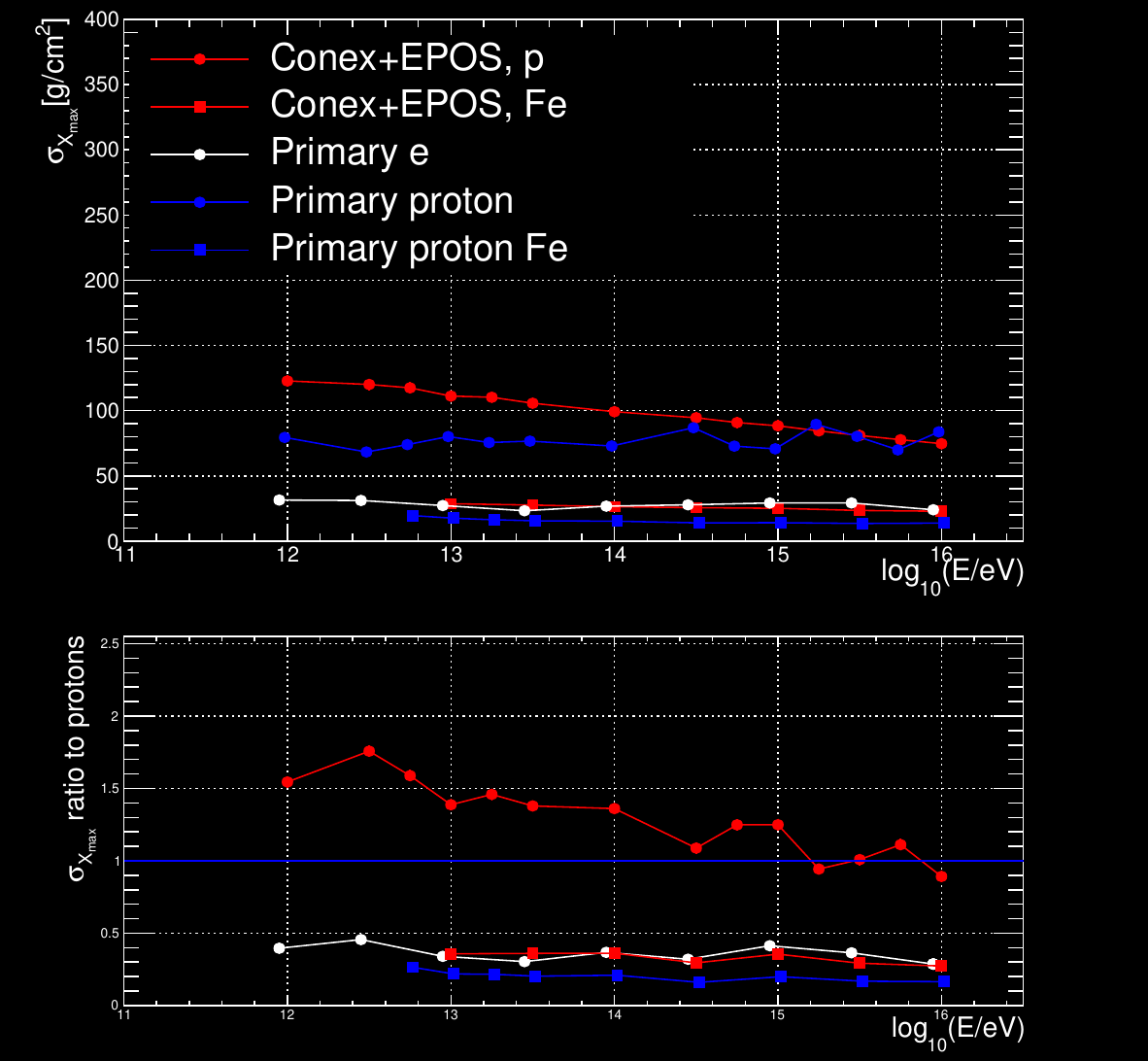}
    \caption{Mean (top) and standard deviation (bottom) of the atmospheric shower maxima as function of \logE{} for the private simulation (blue) and the \Conex{} generator used with the EPOS hadronization model (red). 
    Also shown (dashed) are the predictions from the simple logarithmic models as in~\cite{MATTHEWS2005387} for both EM and hadronic showers, the latter also with an additional constant shift of 100~\gcm{} leading to a better agreement.
     }
    \label{fig:std}
\end{figure}

In order to find the best parameters, we performed a qualitative comparison 
to the \Conex{} generator~\cite{Bergmann:2006yz} using SIBYLL~\cite{Riehn:2019jet} (version 2.3) or EPOS~\cite{Pierog:2009zt} (the LHC tune) as hadronic interaction models.

Figure~\ref{airsim_cmp_SIBYLL} in Appendix~\ref{app:tuning} shows the comparison of the \Xmax{} between the air showers generated according to the model presented in this paper with varied parameters describing the interactions to the \Conex{} generator using EPOS. A similar agreement is found using the SIBYLL model.

Modifying the parameter $C$ to smaller values (\emph{i.e.} less charged hadrons produced per collision) leads to deeper showers, with \Xmax{} closer to that predicted by \Conex{}. However, low values of $C \approx 2$ are not realistic and therefore, similarly to \cite{MATTHEWS2005387}, we fix $C$ to 10 with fluctuations using integers drawn from a Gaussian of a width of $3$.
As for the rest of the parameters we choose the following values: inelasticity of 0.45 with a Gaussian width of 0.2, EM particle flight length maximally $k_\mathrm{EM} X_0$ with $k_\mathrm{EM} = 1.125$, and effectively no limit on the hadronic interaction length of particles within the shower. This allows deeper showers and enhances multimodal longitudinal profiles.

The selected model performance w.r.t. \Conex{} is shown in Figure~\ref{fig:std}, comparing both the \Xmax{} and \sigmaX{} of the presented model for proton as well as iron-initiated showers, and for electromagnetic showers. Also shown are the logarithmic predictions for \Xmax{} as presented in~\cite{MATTHEWS2005387}.

Hadronic showers still exhibit smaller \Xmax{} than \Conex{}, an observation similar to that in~\cite{MATTHEWS2005387}, by about 100~\gcm{}.
As for the \Xmax{} standard deviation, the model predicts roughly a constant width of 75~\gcm{} while \Conex{} predict about twice this value at \logE{} of 12, with a linear decrease to 75~\gcm{} at \logE{} of 16.
The presented model reproduces the well-known smaller \sigmaX{}
for ion-initiated showers, here illustrated on the case of iron (square points in Figure~\ref{fig:std}). While the model cannot substitute the wealth of the hadronic interactions, an indication being the different width of the \Xmax{} distribution, ratios could still be a measure of comparison between different model settings and processes incorporated.

The important internal property of the model is the fact that hadrons energies are not simply of the same value of $E/N$ for each hadron produced in each interaction but rather a random choice from a distribution of a form $\sim x (x_0 - x) \, \exp(-x)$, allowing for a smooth energy distribution among the products, leading to more realistic and random shower profiles, and also allowing resonant conditions to take place for the case of possible new physics resonance models.

The tuned EM model has the \Xmax{} energy dependence close to the simple logarithmic prediction and can be considered as an effective model of more realistic shower fluctuations incorporating the fact that in reality, many more softer photons are radiated along the electron path,  similarly for the $\gamma$ conversion process.

For each energy point we generated 50 events for tuning, 5k events for the selected central model, and 1--10k events for the case of new physics modification (see the next Section) except for few highest energy bins where, due to longer CPU times, we generate close to 1k events. We generated 5--10k \Conex{} showers depending on the energy.


\section{Modification for new physics}

In order to test the possible shower profile and depth modification due to a hypothetical new physics process, we employ the following simple model.
Noticing that the resonant fixed-target setup energy of an incident particle hitting a proton at rest to tune to the mass of a resonance $M_X$ is
$$E_\mathrm{res} \doteq \frac{M^2_{X}}{2m_p}$$
we design a resonance model within the reach of the simulation, namely of masses of 100~GeV and 1~TeV and of decay width of 10\% of its mass. Although large classes of such models are explicitly excluded by direct searches at colliders, we study this scenario as an example scalable to higher energies. We are further motivated by the fact that fluctuations in the \Xmax{} as well as \sigmaX{} spectrum are actually observed, see \emph{e.g.}~\cite{Boncioli:2025flr}, although they can be attributed to statistical fluctuations. Also, a 0.5--1 TeV particle has been recently proposed as a possible model of galactic halo excess in the gamma rays~\cite{Totani:2025fxx}.

The resonant energy of the incident proton hitting a proton at rest corresponding to the production of a resonance of 100~GeV is 5.37~TeV, or, varying by the assumed width and in terms of the \logE{}, in the range of $[12.64, 12.81]$, the central value being 12.73.
For the 1\% width case the interval is $[12.71, 12.74]$, \emph{i.e.} not containing the 12.75 point used, therefore we also simulate a 3\% width resonance with the resonant \logE{} interval of $[12.700, 12.752]$. The 12.75 point will be highly sensitive to the resonance at the onset of the shower.

For primary particles of energies above the resonance production threshold it takes several hundreds of particles be produced to reach the resonance region at the splitting level of roughly 2--5, which allows for a possible observation of variations in the later shower profile.
This is allowed by the smooth energy fractions of produced hadrons within the model. The resonance occurrence within the shower will be sensitive to its width.

For a 1~TeV resonance, the resonant energy is 533~TeV and the range within its width of 100~GeV is $[14.64, 14.81]$ in terms of \logE{}, with central  value of 14.73.  For 1\% width the interval is $[14.71, 14.74]$. Similarly, we also simulate the 3\% width corresponding to the resonant interval of $[14.700, 14.752]$. Resonance energies for more masses $M_X$ are listed in Table~\ref{tab:Ethrs}.

\begin{table}[]
    \centering
    {\scriptsize 
    \begin{tabular}{l|cc|cc|cc|cc}
$M_X \,[\mathrm{TeV}]$ & 0.1 & 0.5 & 1 & 5 & 10 & 50 & 100 & 500 \\ \hline
$E_\mathrm{res}$ & 5.33 TeV & 133 TeV & 533 TeV & 13.3 PeV & 53.3 PeV & 1.33 EeV & 5.33 EeV & 133 EeV \\
$\log_{10} E_\mathrm{res}/\mathrm{eV}$ & 12.73 & 14.12 & 14.73 & 16.12 & 16.73 & 18.12 & 18.73 & 20.12 \\
\end{tabular}
}
    \caption{Resonant energies $E_\mathrm{res}$ for a primary particle to produce a resonance of mass $M_X$ in an interaction when hitting a proton at rest.}
    \label{tab:Ethrs}
\end{table}

We thus explore two resonance models of masses 100~GeV and 1~TeV with decay widths of 10\% and 3\% relative to their mass.
Going to the extreme case, we only allow the resonance production process to take place within the energy resonance interval, turning other hadronic processes off. We than assume three exclusive decay modes of such a resonance separately, namely, to an electron-positron pair, to a pair of muons, or to a pair of charged pions, with a random energy share of the resonance energy within its decay products.

Next, we evaluate the effects for the presence of these models on the average shower maximum. We compare the \Xmax{} energy dependence of the presented private model with the new physics modification to the one without such a process over the energies ranging in terms of \logE{} from 12 to 16. For a comparison, we also add the \Xmax{} dependence for electron-initiated showers, exhibiting higher shower development maxima, and that corresponding to iron-initiated showers.

\section{Results}

\begin{figure}[p]
    \centering
    \includegraphics[width=1\linewidth]{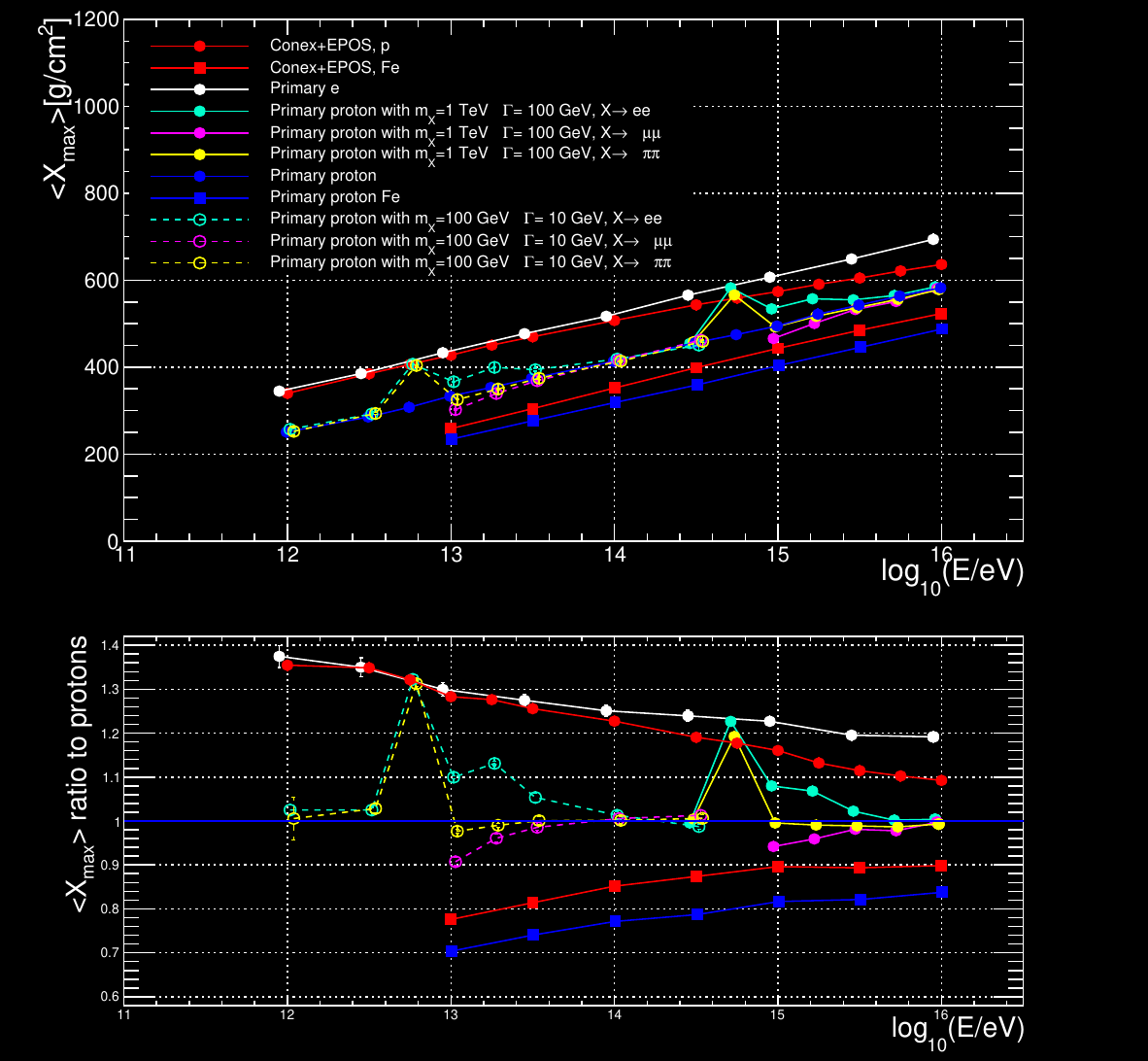} \\
    \caption{Top: Comparison of the \Xmax{} mean values as function of the  logarithm of the initial particle energy for \Conex{}+EPOS showers (red), private simulation described in this study (blue), and for the private shower modified for the presence of a new particle resonance of mass 100~GeV and width 10~GeV  and  mass 1~TeV and width 10~GeV; 
    decaying to a pair of muons (magenta), pions (yellow), or electrons (cyan). For reference, the \Xmax{} profile for a purely electromagnetic shower is also shown (white). Bottom: ratio of the various models to the presented proton model.}
    \label{fig:Zprime_cmp}
\end{figure}

As can be seen in Figure~\ref{fig:Zprime_cmp}, different decay modes of the hypothetical resonance have different impacts on the change in the average \Xmax{} and different sign and length of the persistence of the effect as function of \logE{}: while the EM decay mode deepens the shower for about a decade in \logE{}, the decay to charged pions pair leads to a similar deepening only close to the threshold energy. The resonance decay to muons effectively diminishes the shower visible energy, as we do not model any muon interactions. The visible part of the shower is therefore of a lower energy, leading to more shallow showers. In fact, we do not present the muon decay channel points at the resonant energy, as this leads, within our model, to just two muons and no shower development at all.

It is clear that actually the electromagnetic or muonic decay modes of possible new physics particles may be of most distinct signature in the analysis of the $X_\mathrm{max}$ variable, although here we are neglecting the hadronic sub-shower due to the proton remnant. The mean \Xmax{} is also sensitive to the chemical composition of the primary particles, see also the reference mean \Xmax{} for shower initiated by iron the same figure. 

An example of the average occurrence of the resonance per shower is shown in Table~\ref{tab:freq_100_10_ee} for the case of the resonance of mass of 100~GeV, width 10\% and the $ee$ decay mode.

In order to quote some more details of the shower total composition, for the 100~GeV resonance of a 10\% width simulated at $\logE = 12.9$ ($E=7.9$~TeV), there are in total about 400k EM particles and 1k hadrons and the shower is of $\Xmax{} \approx 400\,$~g/cm${}^2$. In contrast, for the case of no resonance, there are 200k EM particles, 2k hadrons, and $\Xmax{} \approx 300$~g/cm${}^2$.

\begin{table}[!h]
    \centering
    \begin{tabular}{l|cc}
    \logE{} &   resonance occurrence & uncertainty \\ \hline
  12         &  0.000 & 0.000 \\
  12.5       &  0.000 & 0.000 \\
  12.75      &  1.000 & 0.000 \\
  13         &  0.532 & 0.016 \\
  13.25      &  0.542 & 0.016 \\
  13.5       &  0.511 & 0.016 \\
  14         &  0.751 & 0.014 \\
  14.5       &  7.35 & $ < 10^{-3}$ \\
  15         &  33.8 & $ < 10^{-3}$ \\
  15.5       &  75.5 & $ < 10^{-4}$ \\
  16         &  137 & $ < 10^{-4}$ \\
    \end{tabular}
    \caption{Average occurrence of the resonance process per shower as function of the \logE{} and its uncertainty for the resonance of mass of 100~GeV, width 10\% and the $ee$ decay mode.}
    \label{tab:freq_100_10_ee}
\end{table}

\begin{figure}[p]
    \centering
    \includegraphics[width=1\linewidth]{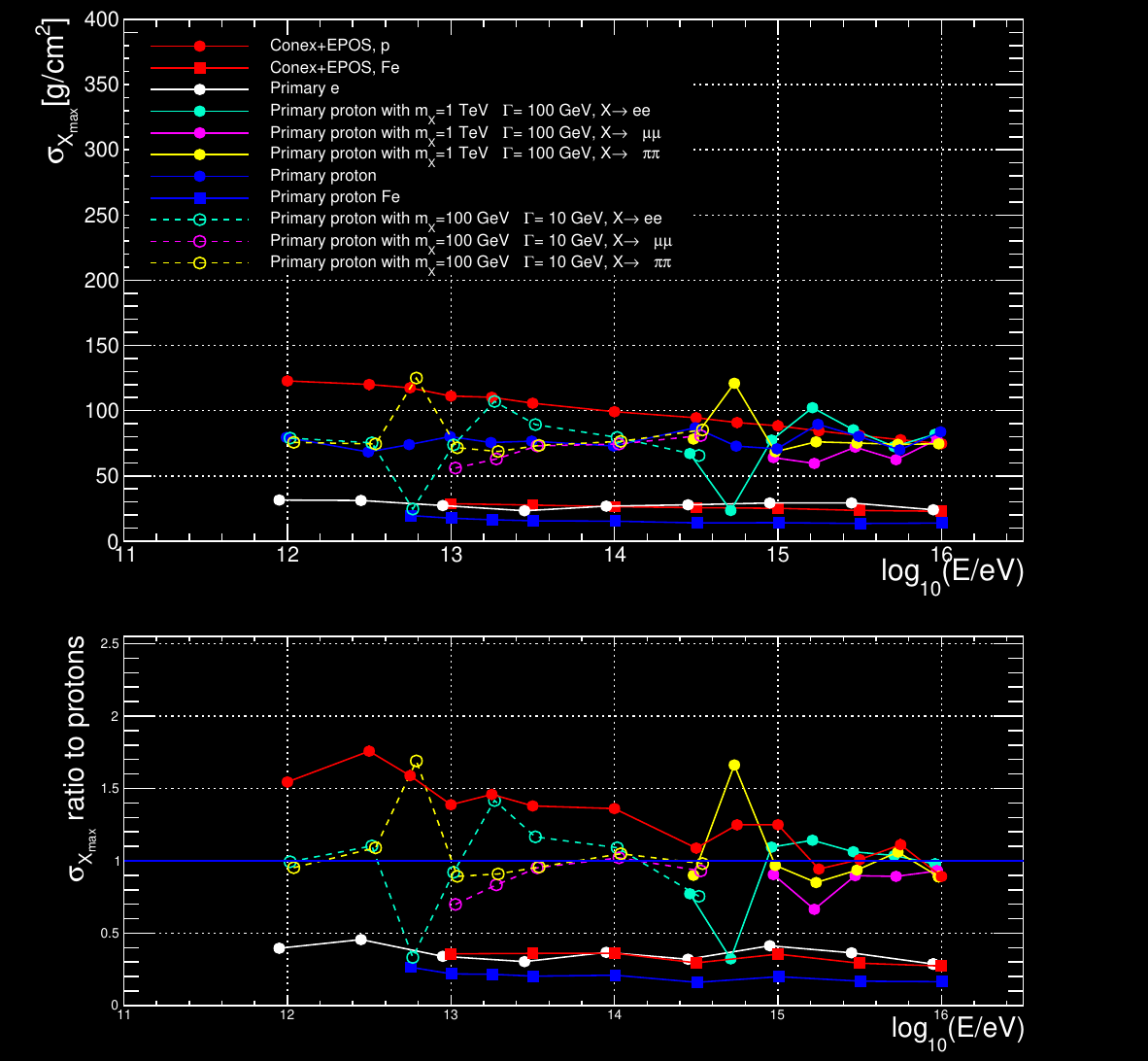} \\
    \caption{Top: Comparison of the standard deviation of \Xmax{} as function of the  logarithm of the initial particle energy for \Conex{}+EPOS showers (red), private simulation described in this study (blue), and for the private shower modified for the presence of a new particle resonance of mass 100~GeV and width 10~GeV  and  mass 1~TeV and width 100~GeV; 
    decaying to a pair of muons (magenta), pions (yellow), or electrons (cyan). For reference, the \sigmaX{} dependence for a purely electromagnetic shower is also shown (white). Bottom: ratio of the various models to the presented proton model.}
    \label{fig:Zprime_cmp_sigma}
\end{figure}

\begin{figure}[!p]
    \centering
    \begin{tabular}{cc}
    \includegraphics[width=0.5\linewidth]{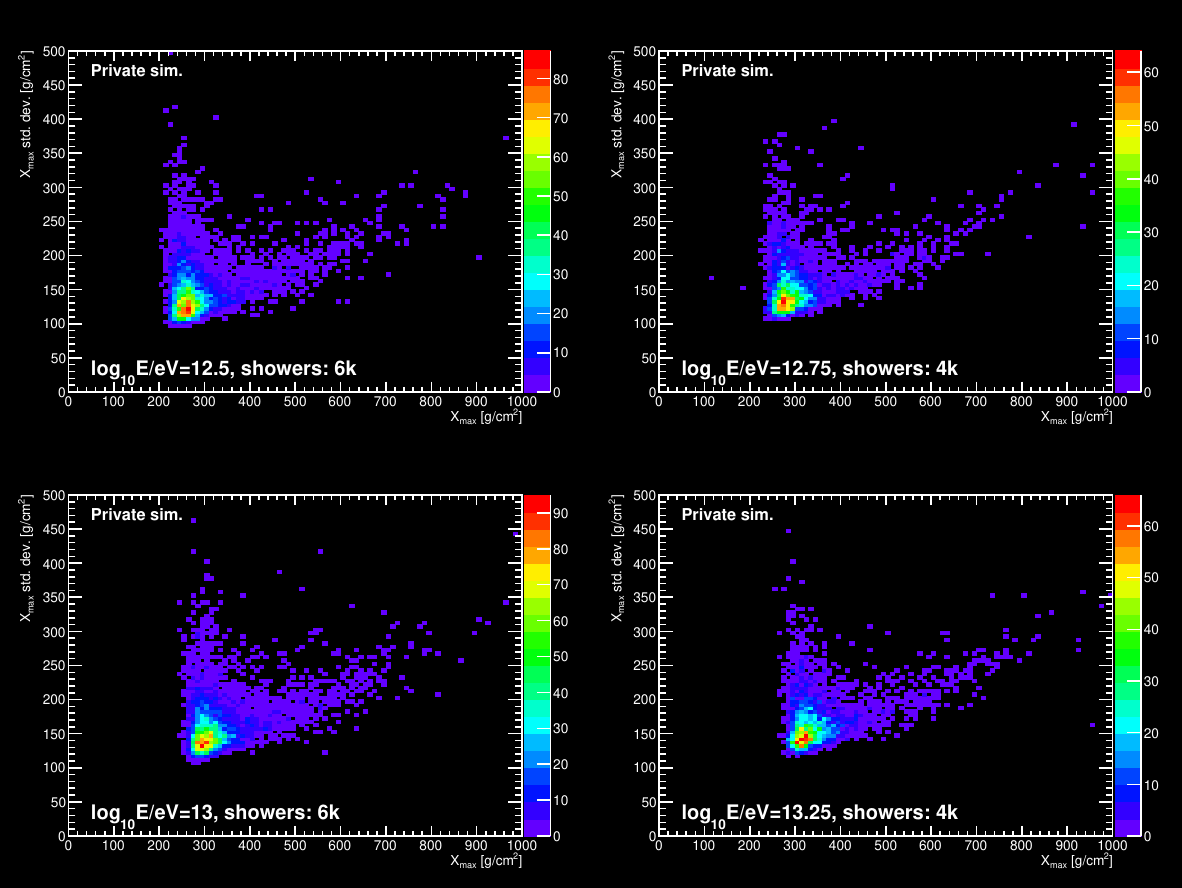} &
    \includegraphics[width=0.425\linewidth] {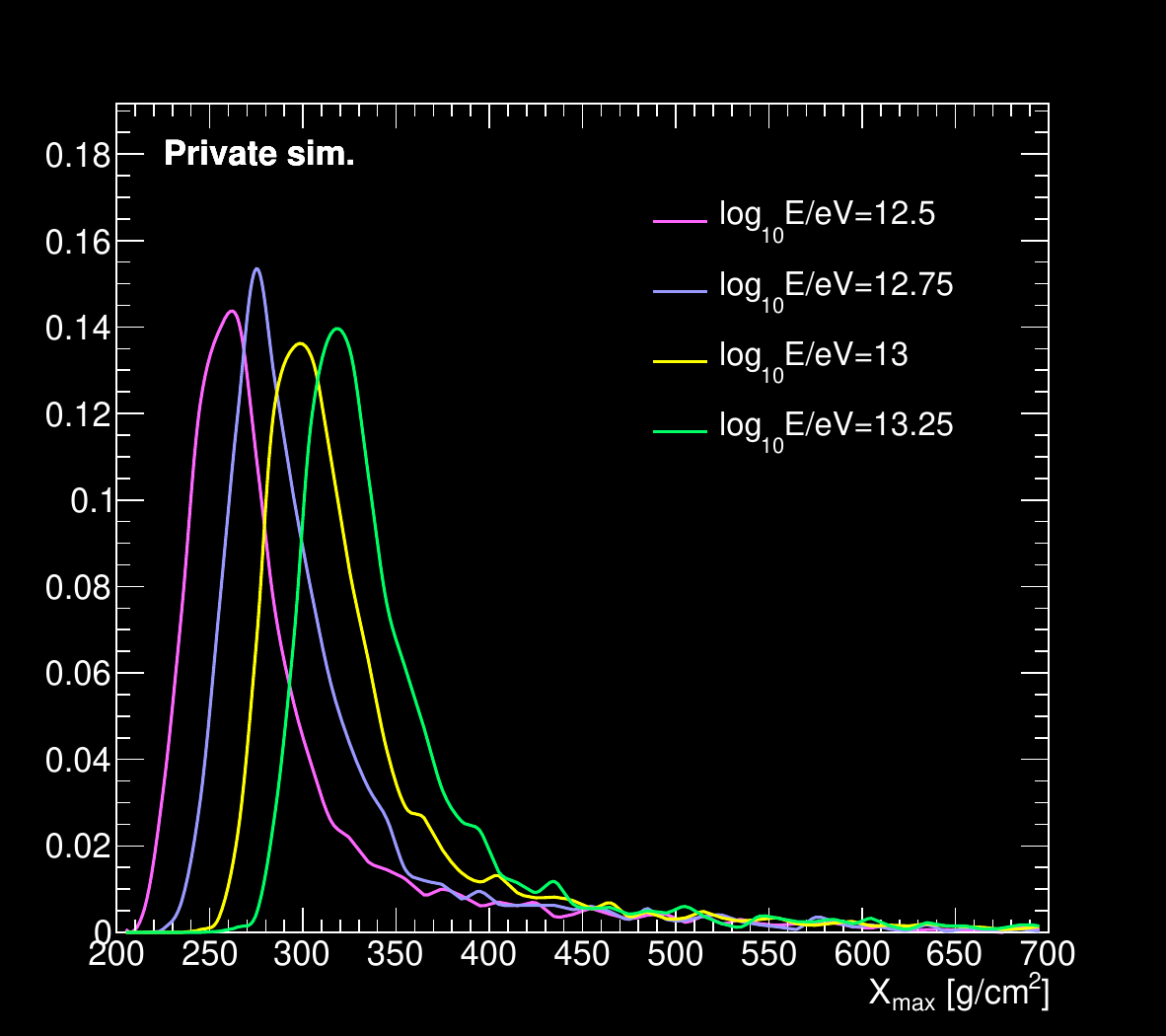} \\
    \includegraphics[width=0.5\linewidth]{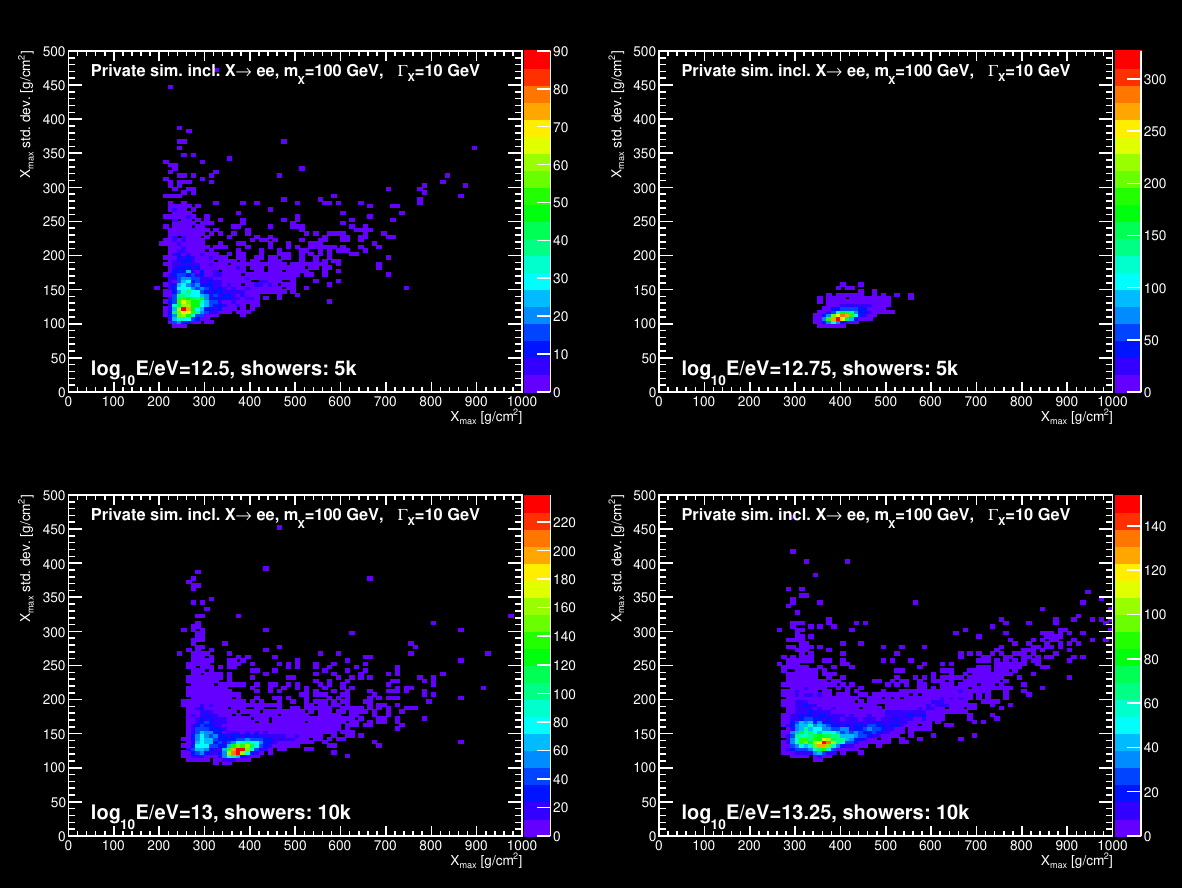} &
    \includegraphics[width=0.425\linewidth] {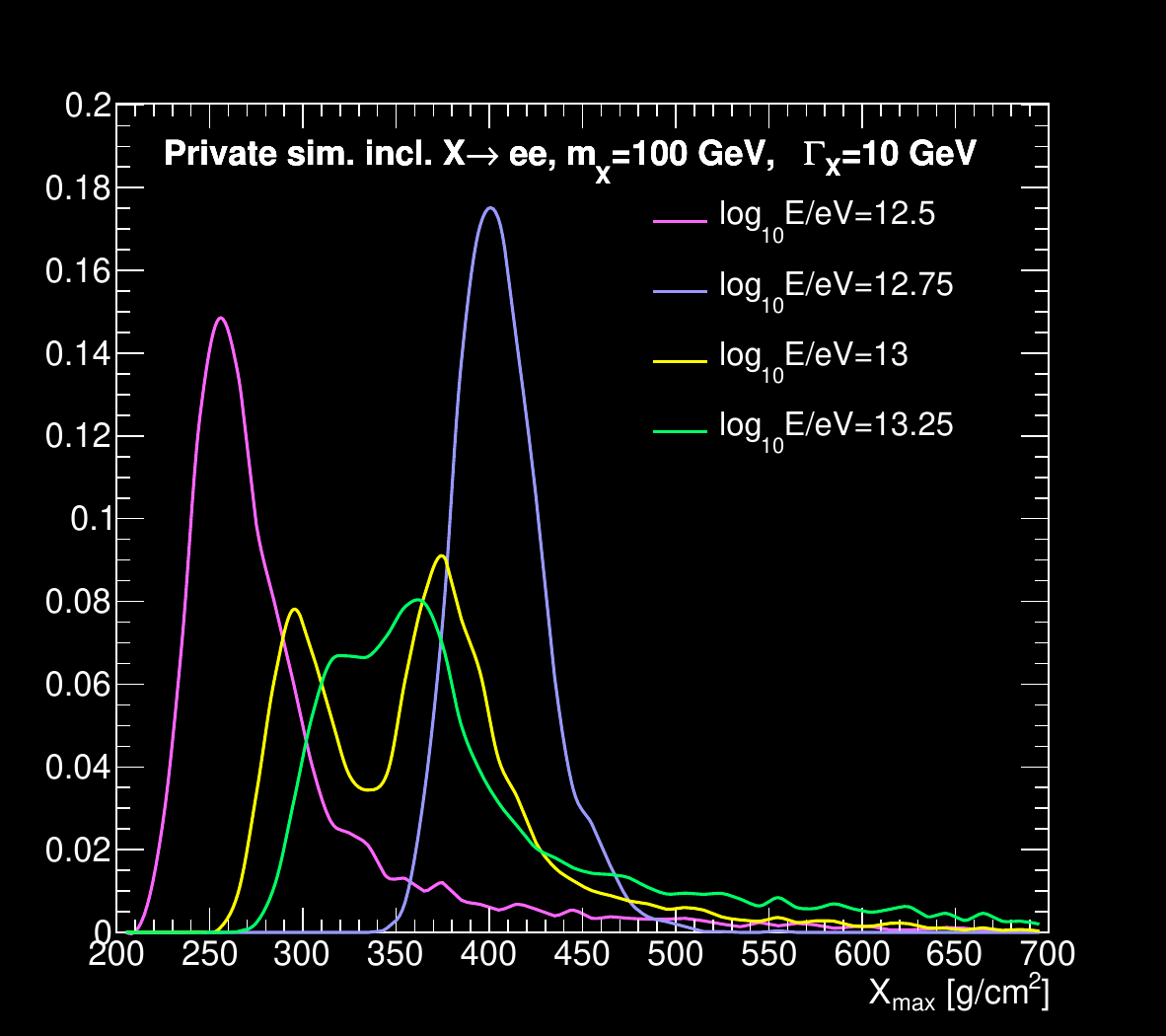} \\
    \includegraphics[width=0.5\linewidth]{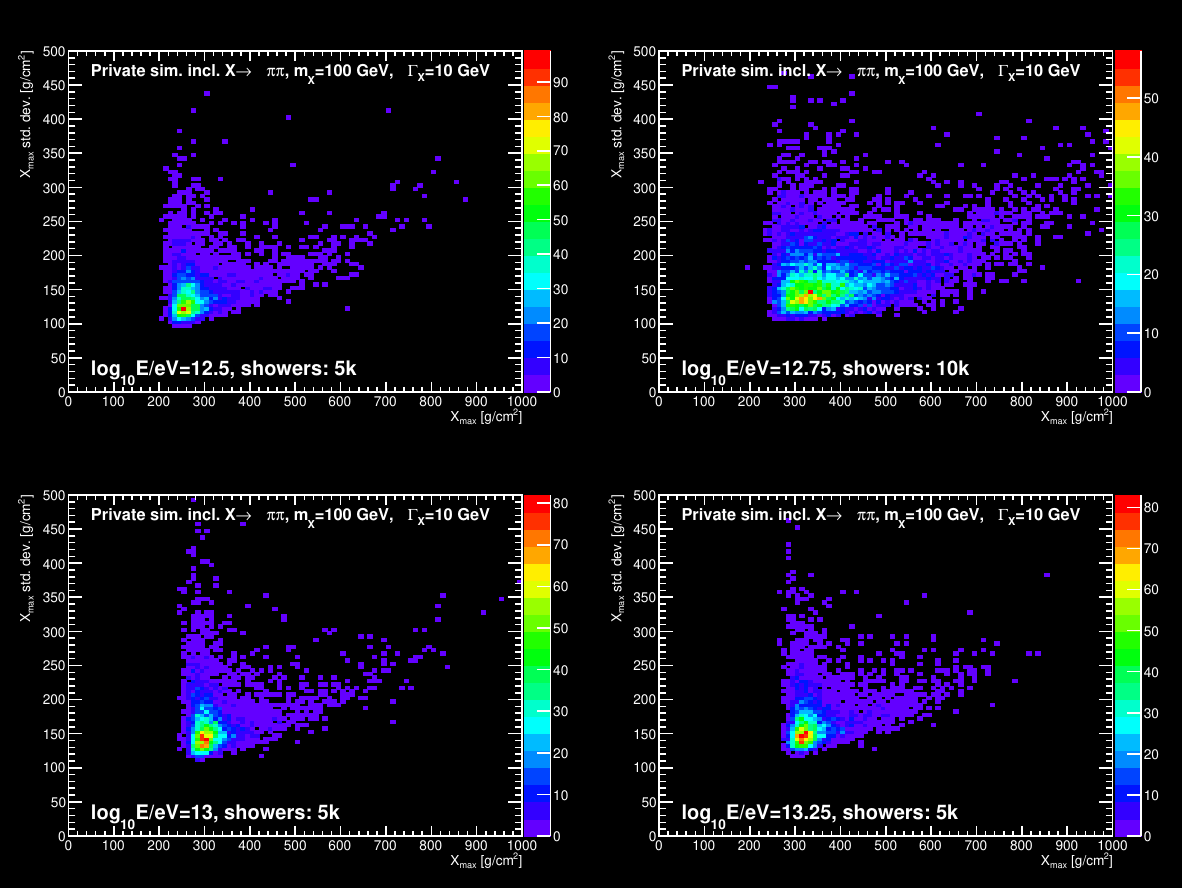} &
    \includegraphics[width=0.425\linewidth] {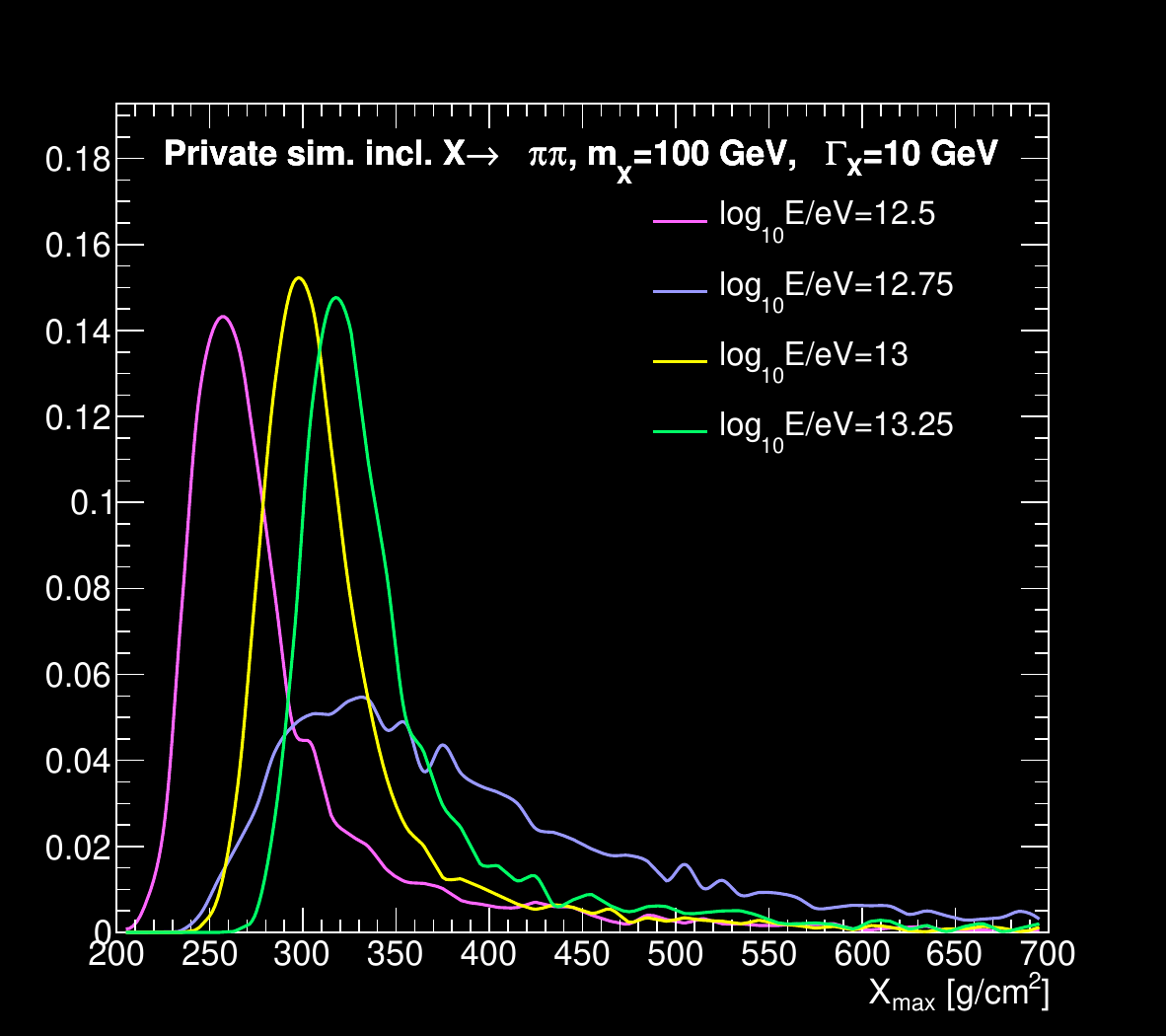} \\
    \end{tabular}
    \caption{Left: Event-by-event comparison of the shower depth standard deviation versus shower \Xmax{} for selected energies. Right: profiles of the \Xmax{} distribution over the showers. 
    Top: The private simulations; middle and bottom: with the additional new physics signal for resonance of mass 100~GeV and width 10~GeV in the $X\rightarrow ee$ and $X\rightarrow \pi\pi$ channels, respectively.
    Structures are visible at or just above the resonant production threshold energies, see text.
    }
    \label{fig:Xmax_2d_cmp}
\end{figure}

Noticing the (anti) correlations in the \Xmax{} and \sigmaX{} depending on the assumed new physics model decay channel, we also take a look at the event-by-event width (computed as the standard deviation) of the shower development plotted against its \Xmax{}, see Figure~\ref{fig:Xmax_2d_cmp}.
This plot shows the squeeze of the shower profile in the $X \rightarrow ee$ channel at the resonance energy threshold of $\logE{} \approx 12.75$  together with the \Xmax{} shift to higher values, and a gradual return of the \Xmax{} profile back to a single maximum with values corresponding to hadronic showers. It is interesting to note the double peak structure above the resonant energy at $\logE{} \approx 13$.
In contrast, in the same figure, the $X \rightarrow \pi\pi$ decay channel exhibits shower broadening possibly due to fluctuations induced in the hadron shower by the leading decay pions at the resonant energy of $\logE{} \approx 12.75$ only.

\begin{figure}[!t]
    \centering
    \begin{tabular}{cc}
    \includegraphics[width=0.5\linewidth]{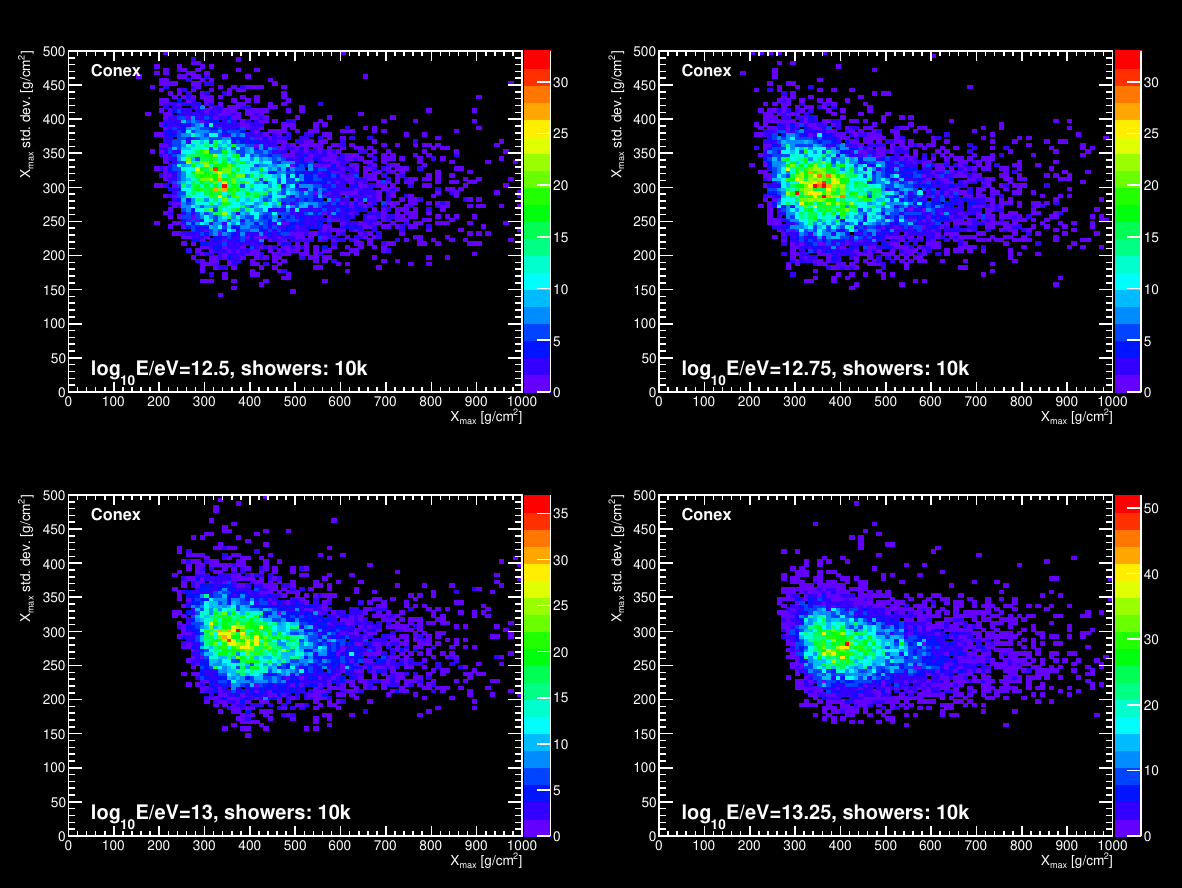} &
    \includegraphics[width=0.425\linewidth] {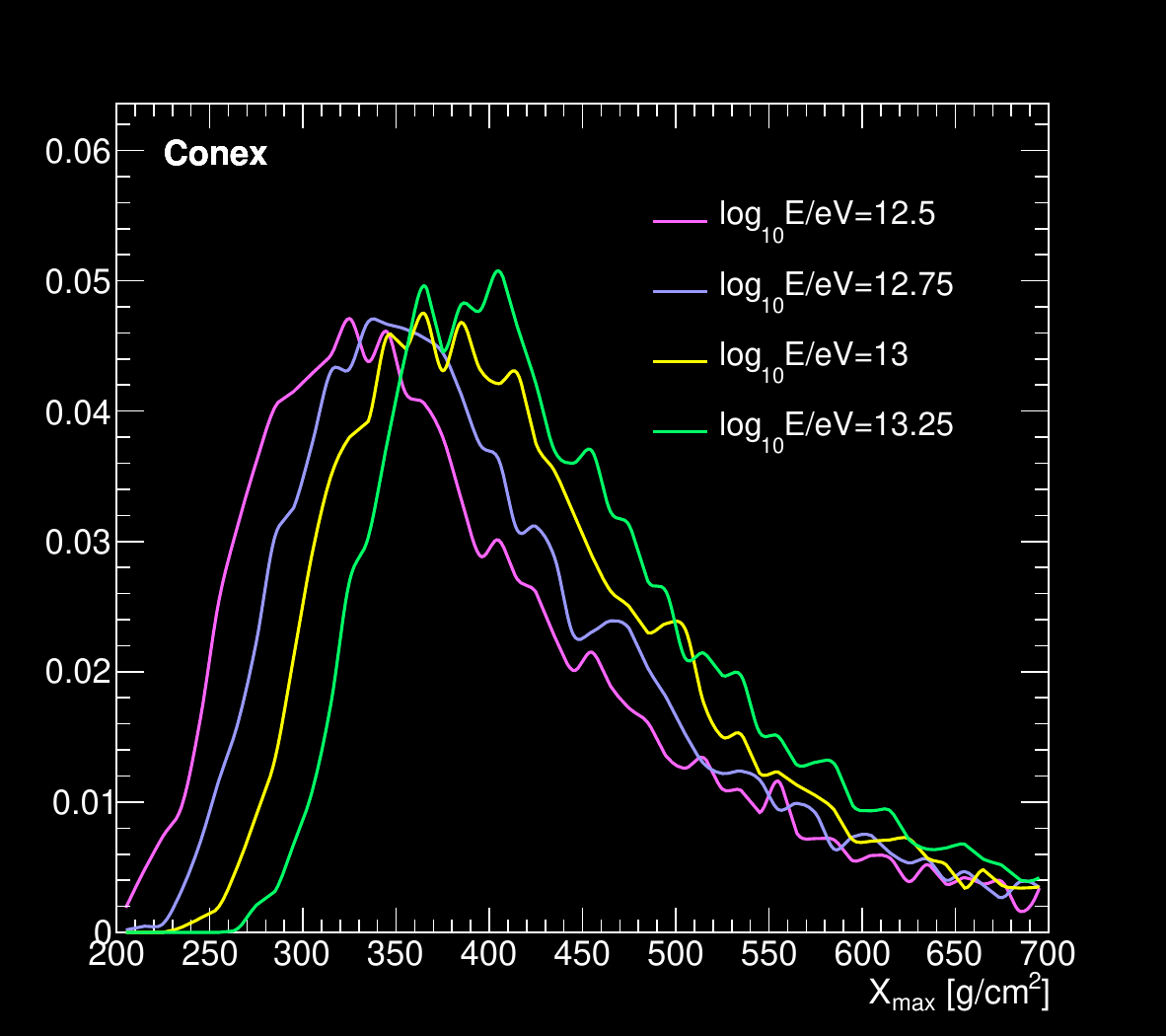} \\
    \end{tabular}
    \caption{Left: Event-by-event comparison of the shower depth standard deviation versus shower \Xmax{} for selected energies. Right: profiles of the \Xmax{} distribution over the showers generated by \Conex{} with EPOS as the hadronization model.
    }
    \label{fig:Xmax_2d_cmp_conex}
\end{figure}

For completeness, it is fair to note that more realistic \Conex{} generator leads to distributions which are about twice broader (see Figure~\ref{fig:Xmax_2d_cmp_conex}), making in reality the possibility to spot the new physics signs more difficult. Nonetheless, even in real data, it may be interesting to study such 2D distributions and search for their broadening or peak structures, explore such variables in more realistic simulations and possibly include them in machine-learning classifiers aimed at determining the type of the primary particle.

\section{Discussion}

The occurrence of the resonance at the beginning of the shower effectively serves as a shower split to two parallel ones, effectively substituting the primary proton by the two decay particles, and making the shower more shallow in a manner similar to that in the superposition model for heavy nuclei. In practice this means there possibly may be an interplay between the new physics and the chemical composition of the primary cosmic rays. The shower depth modification also depends on the resonance dominant decay mode.

These are extreme cases where the resonance production is the dominant process at the resonant energy, a case designed in order to see the direction and shape of the effect. In reality,  more subtle effects are to be expected due to a smaller ratio of the cross-sections of the resonance production to competing SM processes. Also, combination of the presented effects may be in place depending on the resonance allowed decay modes and their partial widths ratios.

The resonance behavior is translated into a threshold-like effect due to the majority of the shower being largely affected only when shortly after the threshold. Still, the effect of the resonance persists for about a decade in \logE{}. The magnitudes of the modified shower depth and its width above the threshold depend strongly on the resonance width (see Appendix~\ref{app:add} for the case of a 3\%-width resonance) due to the fact that a narrower resonance has lower chances to occur within the shower as (within this model) only roughly 10 charged hadrons and the leading proton compete in meeting the resonant energy window.

More detailed morphology of the modification of \Xmax{} is possible to be studied with full simulations.
Yet, the presented model captures the essentials of the behavior of different decay channels, including effectively the one of a resonance decaying into sterile or almost non-interacting particles (here modeled by the non-interacting muons) like neutrinos or some particles beyond the Standard Model.

More importantly, showers with the electromagnetically decaying resonance at the onset of the shower are experimentally indistinguishable from Cherenkov-dominated gamma showers; these are often rejected by primary-nuclei-oriented large observatories, or observed by dedicated gamma observatories.  Conversely, the latter could observe enhanced EM showers at sharp energies regardless of their direction, an indication of such a resonance production decaying electromagnetically, although induced by a primary hadron, success of spotting the process being subject to the experimental energy resolution, EM/hadron showers separation power and depending on the width of the resonance.

Looking at individual shower profiles in terms of both the shower maximum as well as the shower width offers yet another way to look for signs of modifications of the processes within the EAS showers.

Also, a dip in the energy spectrum of hadron-initiated showers would be another hint of the resonance, the number of observed showers at the resonance threshold and higher energies being depleted by the described mechanisms.

\section{Conclusions}

Extending the simple binary energy-even splitting of particles each typical interaction length, we develop a model that allows per-collision fluctuations in particle production and which realistically captures basic shower properties like the shower shape, energy and primary-particle type dependence, reasonably the shower depth and roughly also its width.

Using this model we show that a hypothetical resonance can be observed in the modifications of the depth and shower maximum width of the atmospheric air showers compared to the nominal case, with a sensitivity depending on the resonance width and decay mode. Structures could also be noticed in event-by-event 2D distributions of the individual shower width plotted against its \Xmax{}. These are in practice accessible only by observing the longitudinal shower profiles using their fluorescence light signal.

The experimental challenges remain in the interplay with the non-trivial chemical composition and experimental resolution in \Xmax{}, \sigmaX{} and energy.
Having presented results for two masses of a hypothetical new physics particle allows for projecting the conclusions also to higher masses and energies.

\section{Acknowledgments}
The authors would like to thank the Czech Science Foundation project GA\v{C}R 23-07110S for the support of this work.

\bibliography{main}{}

@article{Pierog:2025ixr,
    author = "Pierog, Tanguy and Werner, Klaus",
    title = "{EPOS LHC-R : a global approach to solve the muon puzzle}",
    eprint = "2508.07105",
    archivePrefix = "arXiv",
    primaryClass = "astro-ph.HE",
    doi = "10.22323/1.501.0358",
    journal = "PoS",
    volume = "ICRC2025",
    pages = "358",
    year = "2025"
}

@article{Boncioli:2025flr,
    author = "Boncioli, Denise",
    collaboration = "Pierre Auger",
    title = "{A new view of UHECRs with the Pierre Auger Observatory}",
    eprint = "2509.15862",
    archivePrefix = "arXiv",
    primaryClass = "astro-ph.HE",
    doi = "10.22323/1.484.0027",
    journal = "PoS",
    volume = "UHECR2024",
    pages = "027",
    year = "2025"
}

@article{Ulrich:2010rg,
    author = "Ulrich, Ralf and Engel, Ralph and Unger, Michael",
    title = "{Hadronic Multiparticle Production at Ultra-High Energies and Extensive Air Showers}",
    eprint = "1010.4310",
    archivePrefix = "arXiv",
    primaryClass = "hep-ph",
    doi = "10.1103/PhysRevD.83.054026",
    journal = "Phys. Rev. D",
    volume = "83",
    pages = "054026",
    year = "2011"
}

@article{PierreAuger:2014ucz,
    author = "Aab, Alexander and others",
    collaboration = "Pierre Auger",
    title = "{Muons in Air Showers at the Pierre Auger Observatory: Mean Number in Highly Inclined Events}",
    eprint = "1408.1421",
    archivePrefix = "arXiv",
    primaryClass = "astro-ph.HE",
    reportNumber = "FERMILAB-PUB-14-290-AD-AE-E-TD",
    doi = "10.1103/PhysRevD.91.032003",
    journal = "Phys. Rev. D",
    volume = "91",
    number = "3",
    pages = "032003",
    year = "2015",
    note = "[Erratum: Phys.Rev.D 91, 059901 (2015)]"
}

@article{Albrecht:2021cxw,
    author = "Albrecht, Johannes and others",
    title = "{The Muon Puzzle in cosmic-ray induced air showers and its connection to the Large Hadron Collider}",
    eprint = "2105.06148",
    archivePrefix = "arXiv",
    primaryClass = "astro-ph.HE",
    doi = "10.1007/s10509-022-04054-5",
    journal = "Astrophys. Space Sci.",
    volume = "367",
    number = "3",
    pages = "27",
    year = "2022"
}

@article{MATTHEWS2005387,
title = "{A Heitler model of extensive air showers}",
journal = {Astroparticle Physics},
volume = {22},
number = {5},
pages = {387-397},
year = {2005},
issn = {0927-6505},
doi = {https://doi.org/10.1016/j.astropartphys.2004.09.003},
url = {https://www.sciencedirect.com/science/article/pii/S0927650504001598},
author = {J. Matthews},
}

@article{Brooijmans:2016lfv,
    author = "Brooijmans, Gustaaf and Schichtel, Peter and Spannowsky, Michael",
    title = "{Cosmic ray air showers from sphalerons}",
    eprint = "1602.00647",
    archivePrefix = "arXiv",
    primaryClass = "hep-ph",
    reportNumber = "IPPP-16-09, DCPT-16-18, MCNET-16-02",
    doi = "10.1016/j.physletb.2016.08.030",
    journal = "Phys. Lett. B",
    volume = "761",
    pages = "213--218",
    year = "2016"
}

@article{Fischer:2020dmn,
    author = "Fischer, Oliver and Reininghaus, Maximilian and Ulrich, Ralf",
    title = "{Avenues to new-physics searches in cosmic ray air showers}",
    eprint = "2012.14293",
    archivePrefix = "arXiv",
    primaryClass = "hep-ph",
    doi = "10.22323/1.390.0602",
    journal = "PoS",
    volume = "ICHEP2020",
    pages = "602",
    year = "2021"
}

@inproceedings{gaisser1977reliability,
  title={Reliability of the method of constant intensity cuts for reconstructing the average development of vertical showers},
  author={Gaisser, Thomas K and Hillas, A Michael},
  booktitle={In: International Cosmic Ray Conference, 15th, Plovdiv, Bulgaria, August 13-26, 1977, Conference Papers. Volume 8.(A79-37301 15-93) Sofia, B\v{u}lgarska Akademiia na Naukite, 1978, p. 353-357. NSF-supported research.},
  volume={8},
  pages={353--357},
  year={1977}
}

@article{Totani:2025fxx,
    author = "Totani, Tomonori",
    title = "{20 GeV halo-like excess of the Galactic diffuse emission and implications for dark matter annihilation}",
    eprint = "2507.07209",
    archivePrefix = "arXiv",
    primaryClass = "astro-ph.HE",
    doi = "10.1088/1475-7516/2025/11/080",
    journal = "JCAP",
    volume = "11",
    pages = "080",
    year = "2025"
}

@article{Riehn:2019jet,
    author = "Riehn, Felix and Engel, Ralph and Fedynitch, Anatoli and Gaisser, Thomas K. and Stanev, Todor",
    title = "{Hadronic interaction model Sibyll 2.3d and extensive air showers}",
    eprint = "1912.03300",
    archivePrefix = "arXiv",
    primaryClass = "hep-ph",
    doi = "10.1103/PhysRevD.102.063002",
    journal = "Phys. Rev. D",
    volume = "102",
    number = "6",
    pages = "063002",
    year = "2020"
}

@article{Pierog:2009zt,
    author = "Pierog, T. and Werner, K.",
    editor = {Capdevielle, Jean-No\"el and Engel, Ralph and Pattison, Bryan},
    title = "{EPOS Model and Ultra High Energy Cosmic Rays}",
    eprint = "0905.1198",
    archivePrefix = "arXiv",
    primaryClass = "hep-ph",
    doi = "10.1016/j.nuclphysbps.2009.09.017",
    journal = "Nucl. Phys. B Proc. Suppl.",
    volume = "196",
    pages = "102--105",
    year = "2009"
}

@article{Bergmann:2006yz,
    author = "Bergmann, Till and Engel, R. and Heck, D. and Kalmykov, N. N. and Ostapchenko, Sergey and Pierog, T. and Thouw, T. and Werner, K.",
    title = "{One-dimensional Hybrid Approach to Extensive Air Shower Simulation}",
    eprint = "astro-ph/0606564",
    archivePrefix = "arXiv",
    doi = "10.1016/j.astropartphys.2006.08.005",
    journal = "Astropart. Phys.",
    volume = "26",
    pages = "420--432",
    year = "2007"
}

@misc{githubjk,
  author       = {Jiri Kvita},
  title        = {AirShowers},
  year         = {2025},
  howpublished = {\url{https://github.com/jirikvita/myStatsToys/tree/main/AirShowers}},
  note         = {github repository}
}
\bibliographystyle{unsrt}

\clearpage
\appendix{}

\section{Additional width of the new physics resonance}
\label{app:add}

\begin{figure}[!h]
    \centering
    \includegraphics[width=1\linewidth]{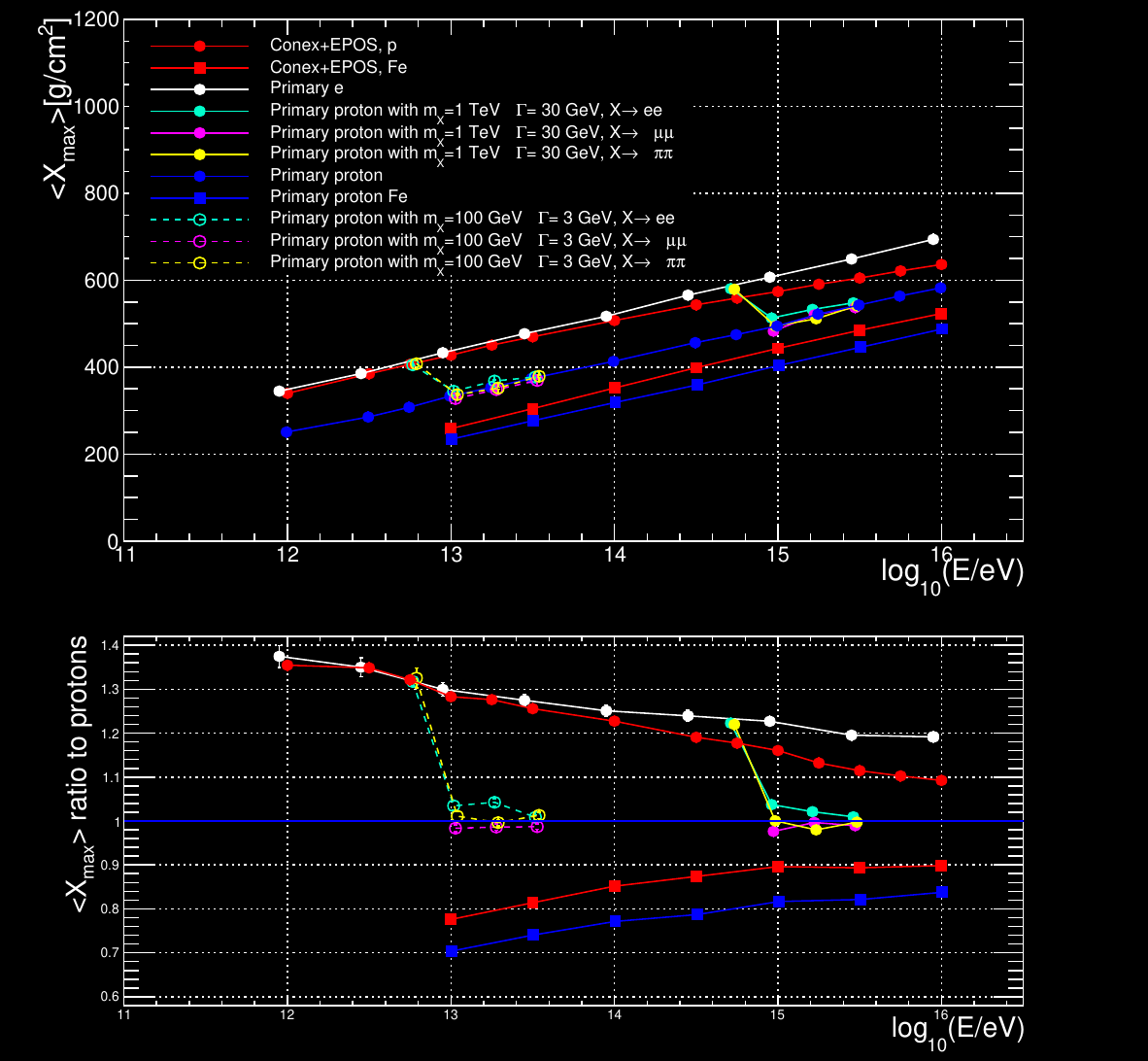} \\
    \caption{Top: Comparison of the \Xmax{} mean values as function of the  logarithm of the initial particle energy for \Conex{}+EPOS showers (red), private simulation described in this study (blue), and for the private shower modified for the presence of a new particle resonances of mass 100~GeV and width 3~GeV and  mass 1~TeV and width 30~GeV; decaying to a pair of muons (magenta), pions (yellow), or electrons (cyan). For reference, the \Xmax{} profile for a purely electromagnetic shower is also shown (white). Bottom: ratio of the various models to the presented proton model.}
    \label{fig:Zprime_cmp_3ptcl}
\end{figure}

\begin{figure}[p]
    \centering
    \includegraphics[width=1\linewidth]{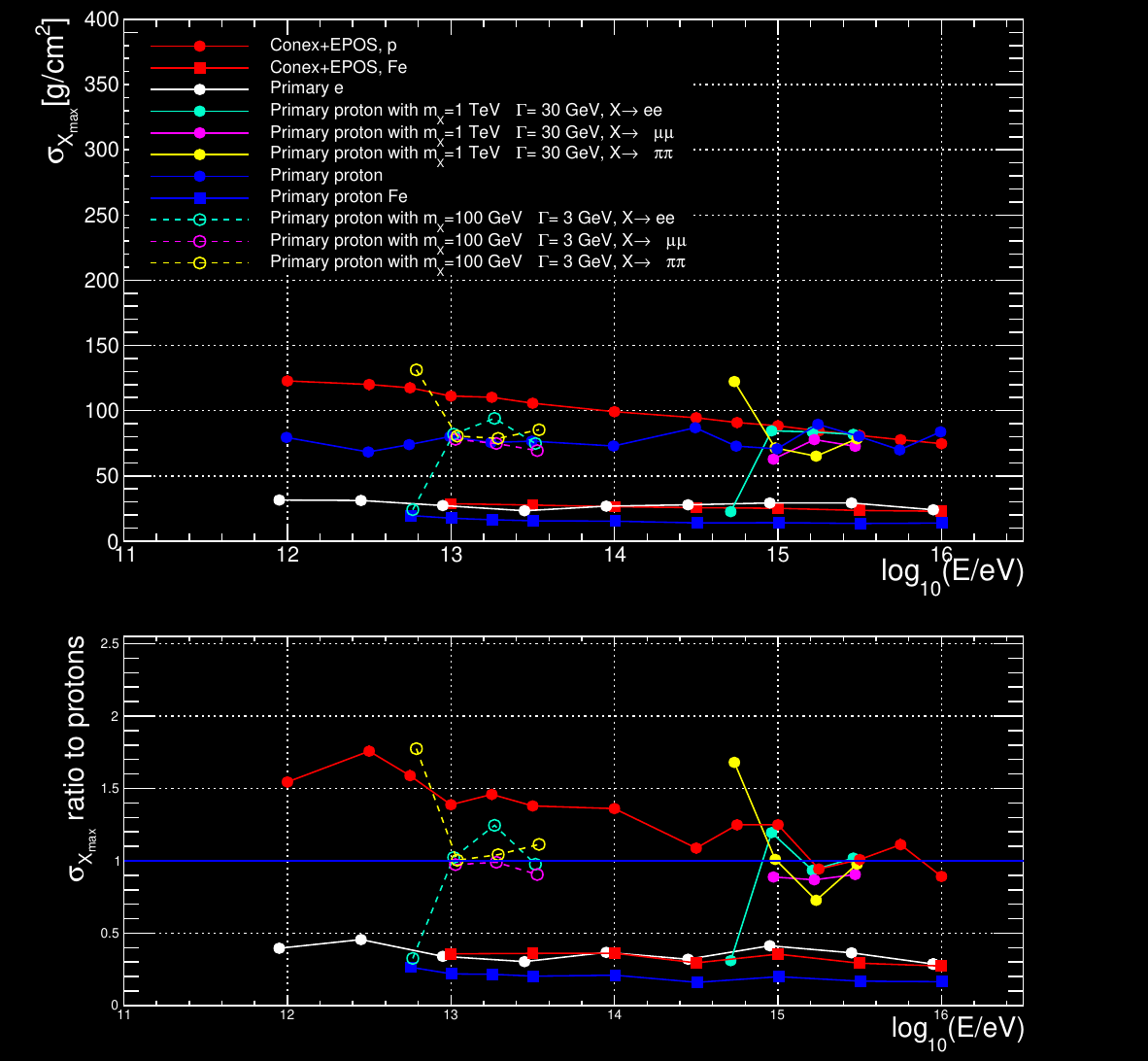} \\
    \caption{Top: Comparison of the standard deviation of \Xmax{} as function of the  logarithm of the initial particle energy for \Conex{}+EPOS showers (red), private simulation described in this study (blue), and for the private shower modified for the presence of a new particle resonance of mass 100~GeV and width 3~GeV  and  mass 1~TeV and width 30~GeV; 
    decaying to a pair of muons (magenta), pions (yellow), or electrons (cyan). For reference, the \sigmaX{} dependence for a purely electromagnetic shower is also shown (white). Bottom: ratio of the various models to the presented proton model.}
    \label{fig:Zprime_cmp_sigma_3ptcl}
\end{figure}

Here we present also a mode with a 3\% decay with of the hypothetical resonance, showing the resulting shower depths in Figure~\ref{fig:Zprime_cmp_3ptcl} and \sigmaX{} in Figure~\ref{fig:Zprime_cmp_sigma_3ptcl}.

\clearpage
\section{Private model shower parameters tuning}
\label{app:tuning}

\begin{figure}[!h]
\begin{tabular}{ccc}
\includegraphics[width=1.0\textwidth]{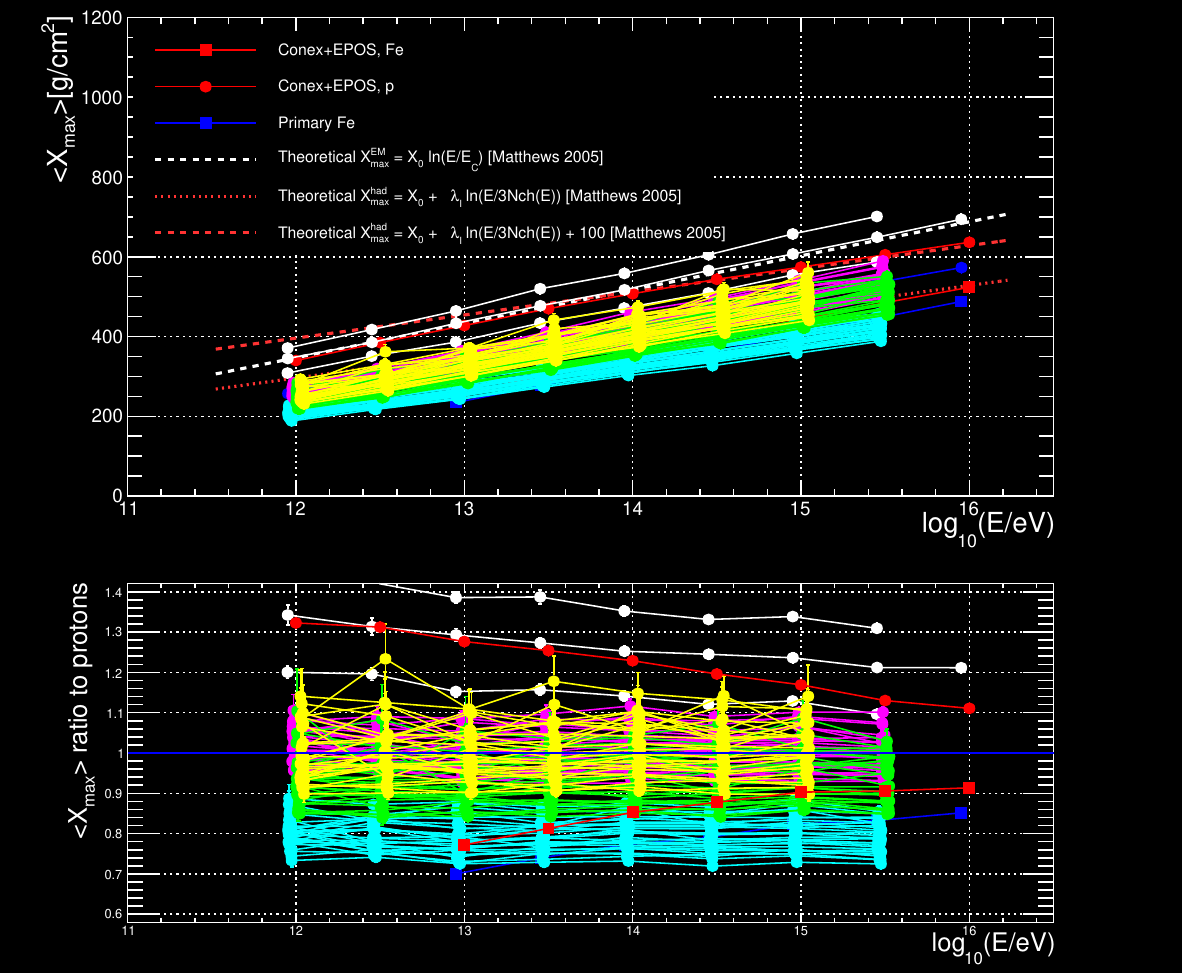} 
\end{tabular}
\caption{Various private shower parameters models compared to the \Conex{} generator with EPOS as the hadronization model. Models in cyan, green, magenta correspond to the maximal EM particle interaction at $k_\mathrm{EM} = 1.0$, $1.125$, $1.5$ as multiples of the process (radiation or pair creation) length, respectively, with various lengths allowed for the hadronic processes and with number of pions produced in range $C=8$--$12$ and inelasticities $0.4$--$0.6$. Models in yellow  correspond to (unphysically) small multiplicities $C=2$--$6$ with maximal EM $k_\mathrm{EM}=1.125$.}
\label{airsim_cmp_SIBYLL}
\end{figure}

We show the overview of shower parameters tuning in classes of the EM shower maximal development per event in terms of the radiation length. Figure~\ref{airsim_cmp_SIBYLL} shows the resulting average \Xmax{} profiles. 

\begin{figure}[!h]
    \centering
    \includegraphics[width=1.0\linewidth]{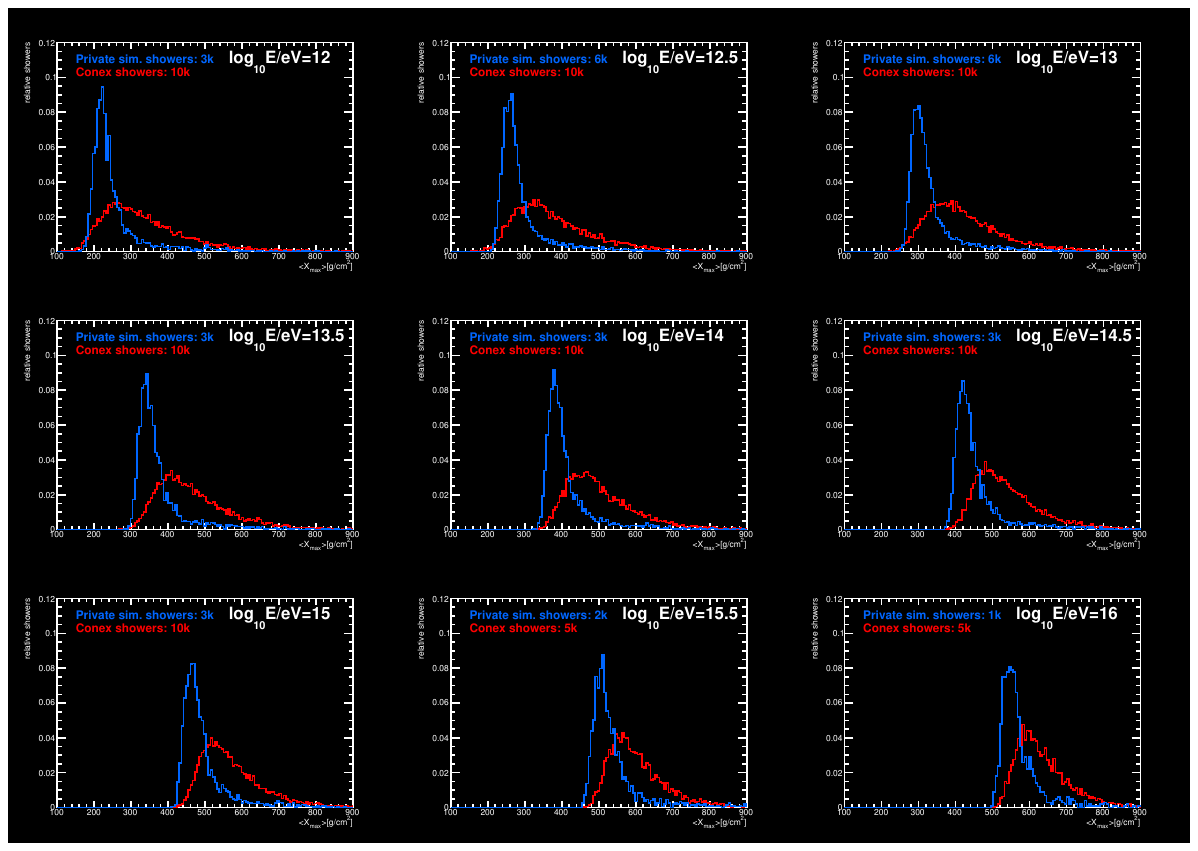} 
    \caption{The distribution of the \Xmax{} for the \Conex{} generator with EPOS as the hadronization model (red) and for the presented private showers (blue).}
    \label{fig:conex_profiles_cmp_epos}
\end{figure}

In Figure~\ref{fig:conex_profiles_cmp_epos} we also show an example of the \sigmaX{} for the central model and the \Conex{} generator.

\clearpage
\section{Multimodal showers}

\begin{figure}[!h]
    \centering
\begin{tabular}{cc}
    \includegraphics[width=0.62\linewidth]{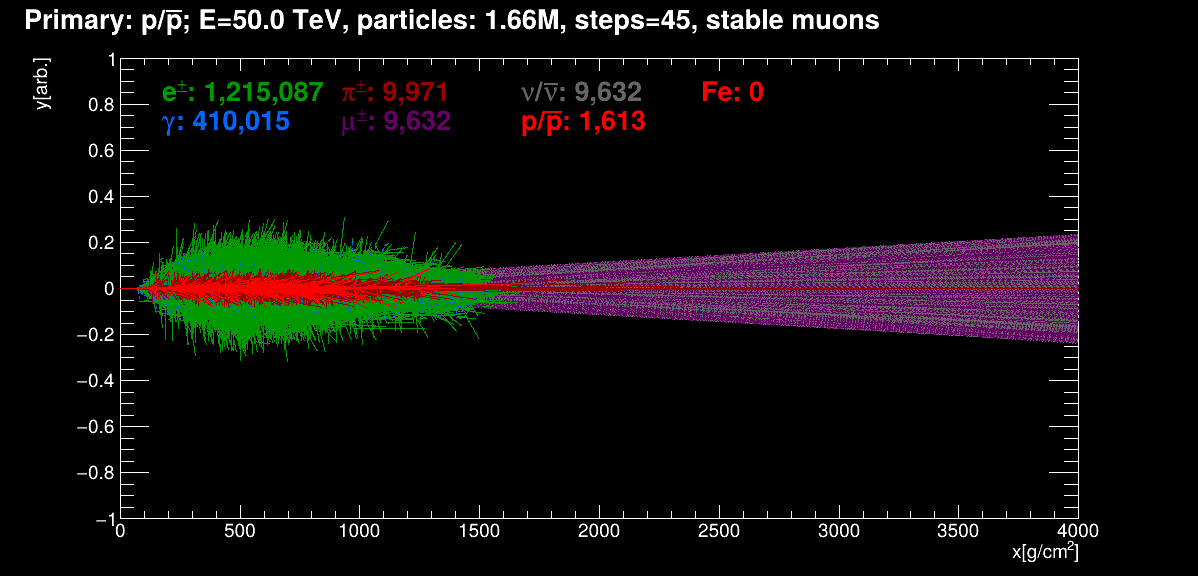}  &   \includegraphics[width=0.36\linewidth]{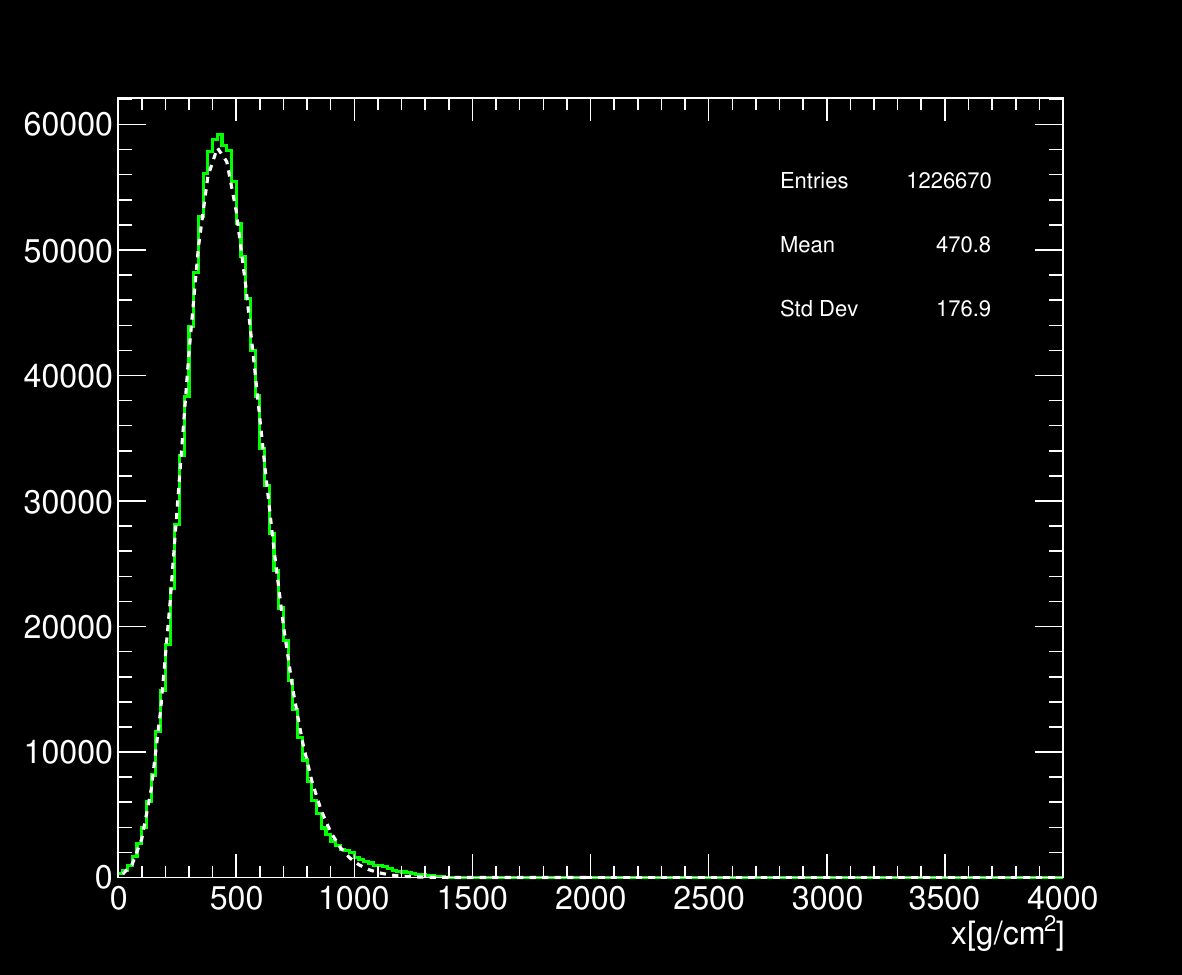} \\
    \includegraphics[width=0.62\linewidth]{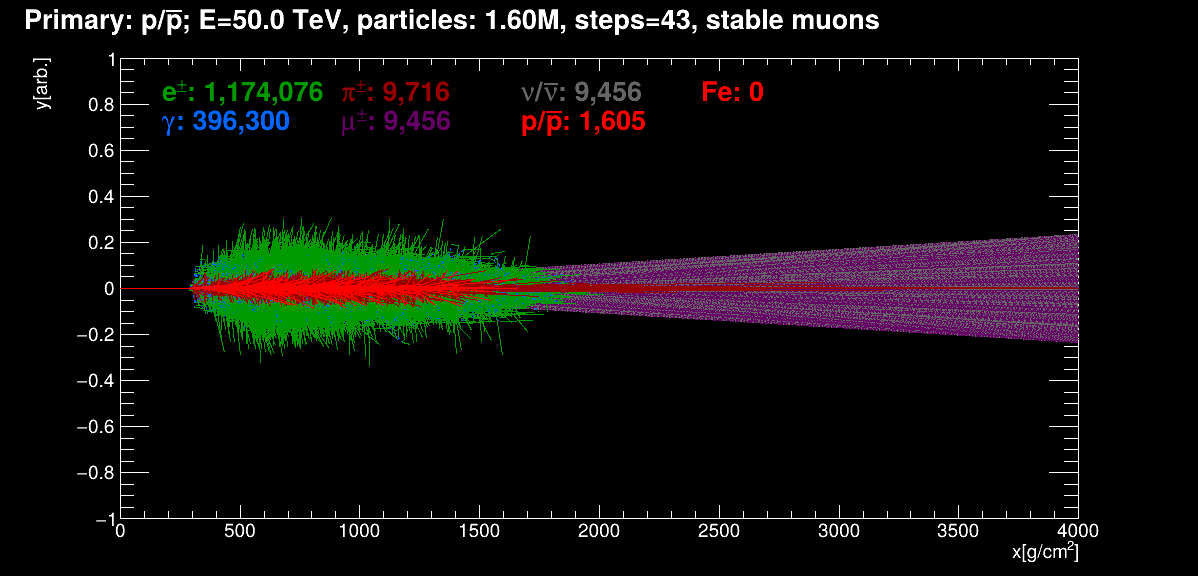} & 
    \includegraphics[width=0.36\linewidth]{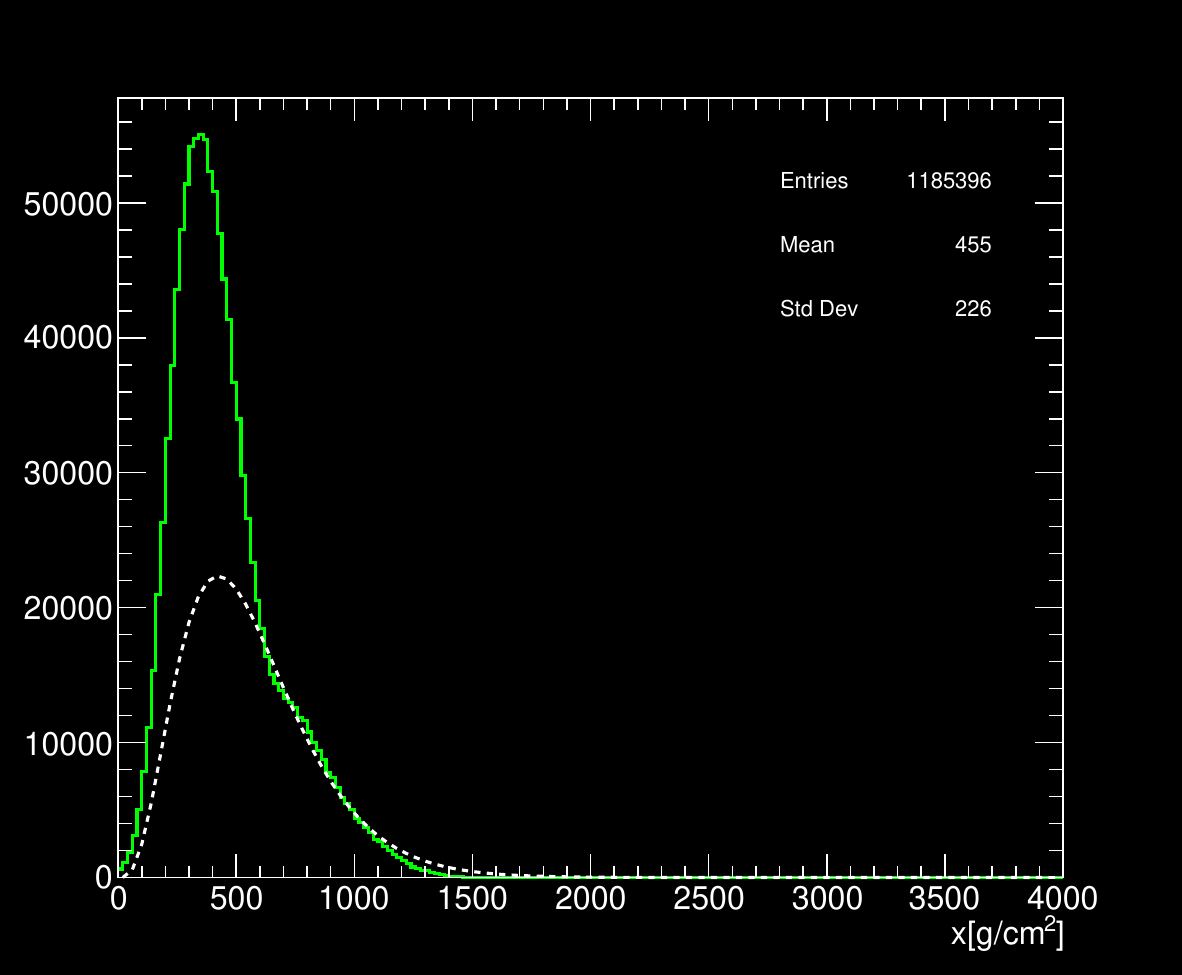} \\
    \includegraphics[width=0.62\linewidth]{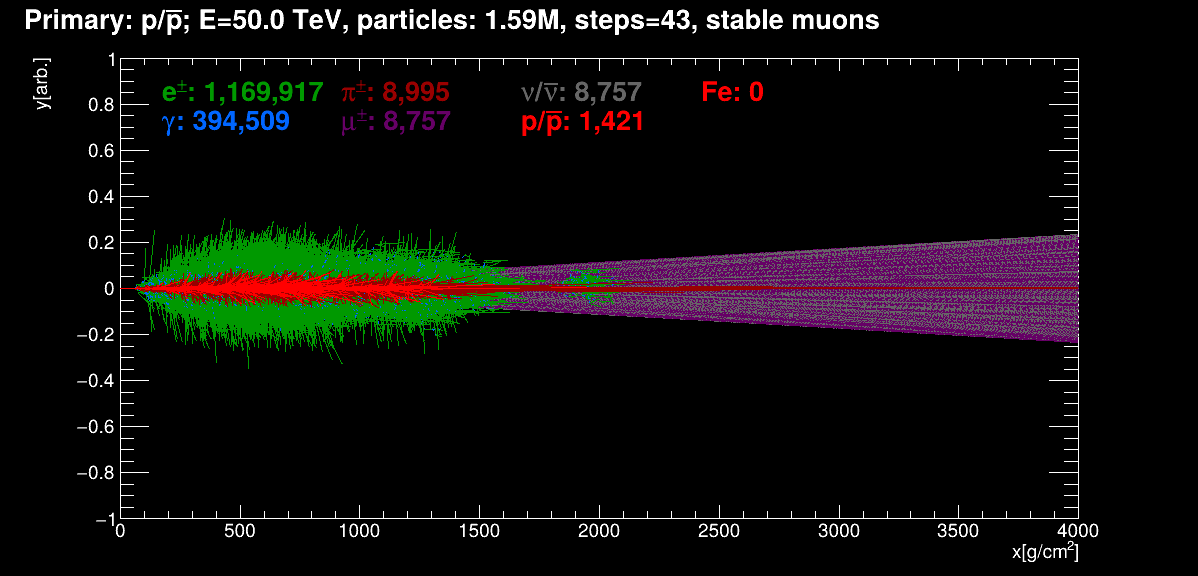} & 
    \includegraphics[width=0.36\linewidth]{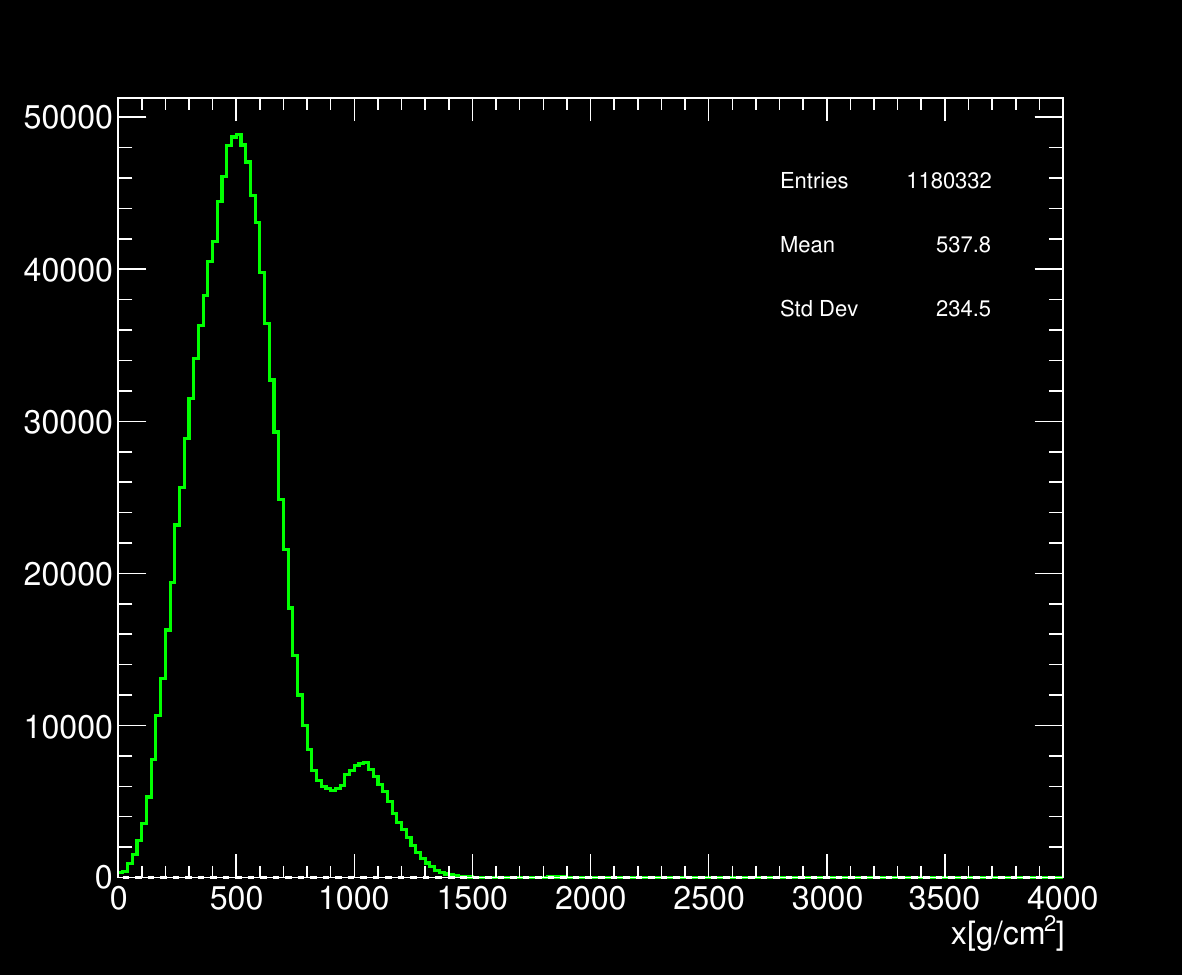} \\
    \includegraphics[width=0.62\linewidth]{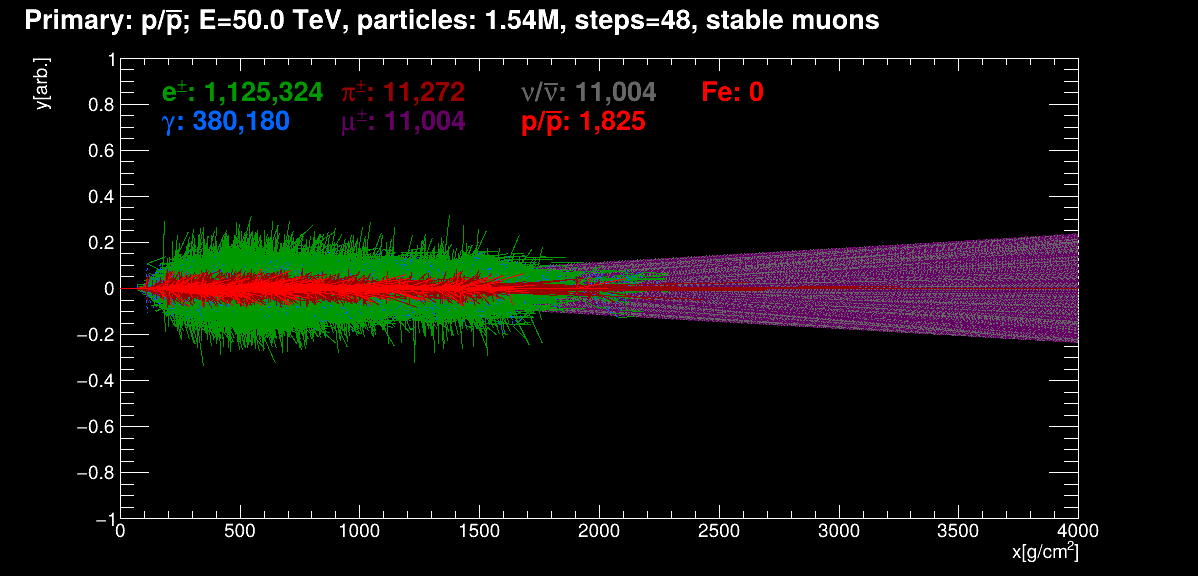} & 
    \includegraphics[width=0.36\linewidth]{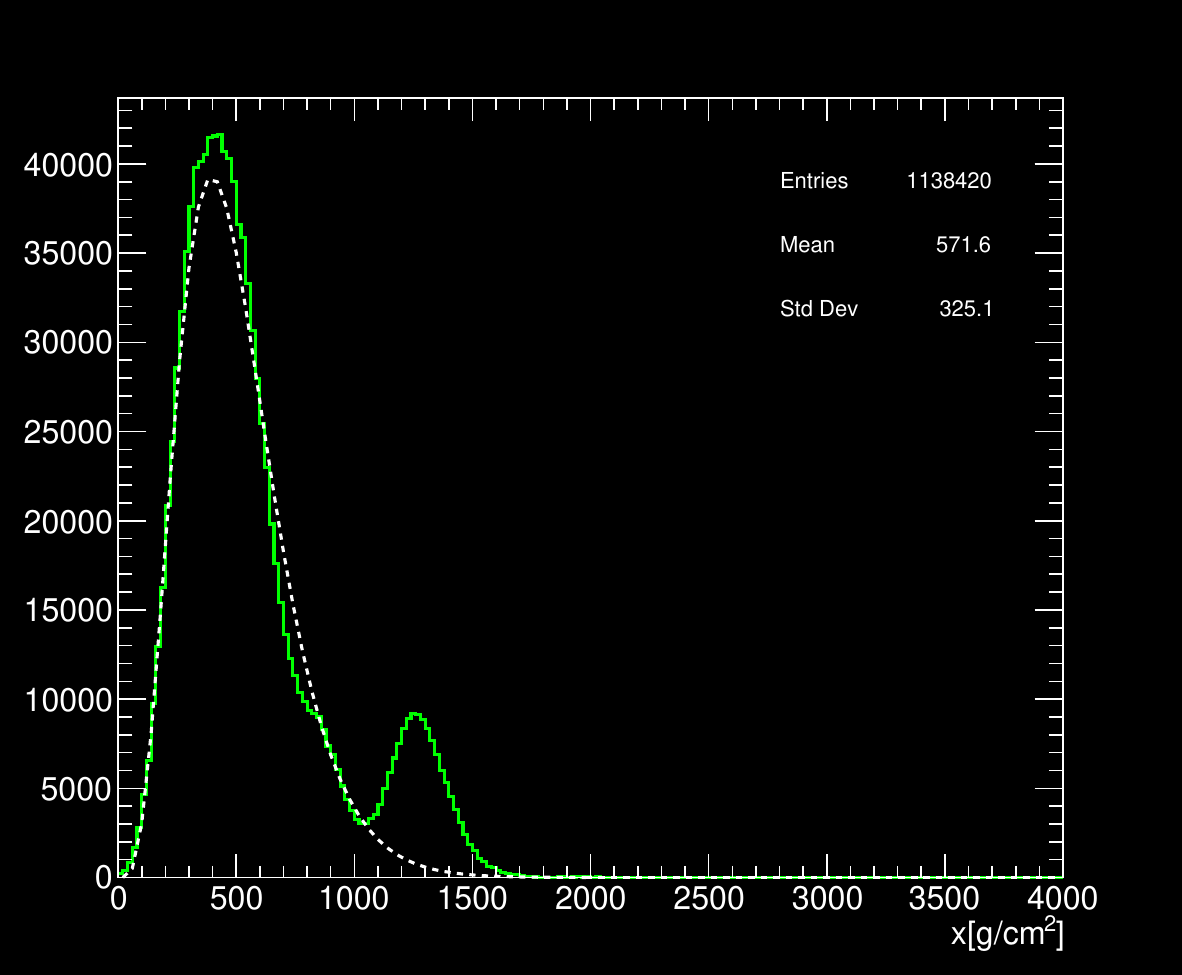}
\end{tabular}
    \caption{A~visualization of private parameterized atmospheric air showers initiated by a~50~TeV proton exhibiting a single (top) and gradually more distinct double or triple peak structure (bottom plots). Right: Longitudinal shower developments in terms of the number of charged particles as function of the atmospheric depth for the two showers.}
    \label{fig:single_double}
\end{figure}

It can be observed that even such a simple model naturally exhibits non-standard shower developments due to the allowed randomness of the shower development, leading to the presence of multiple maxima in the shower longitudinal profile, see Figure~\ref{fig:single_double} as an example of the possible structures of a gradual appearance of additional local maxima.

\begin{figure}[!h]
    \centering
    \includegraphics[width=1.0\linewidth]{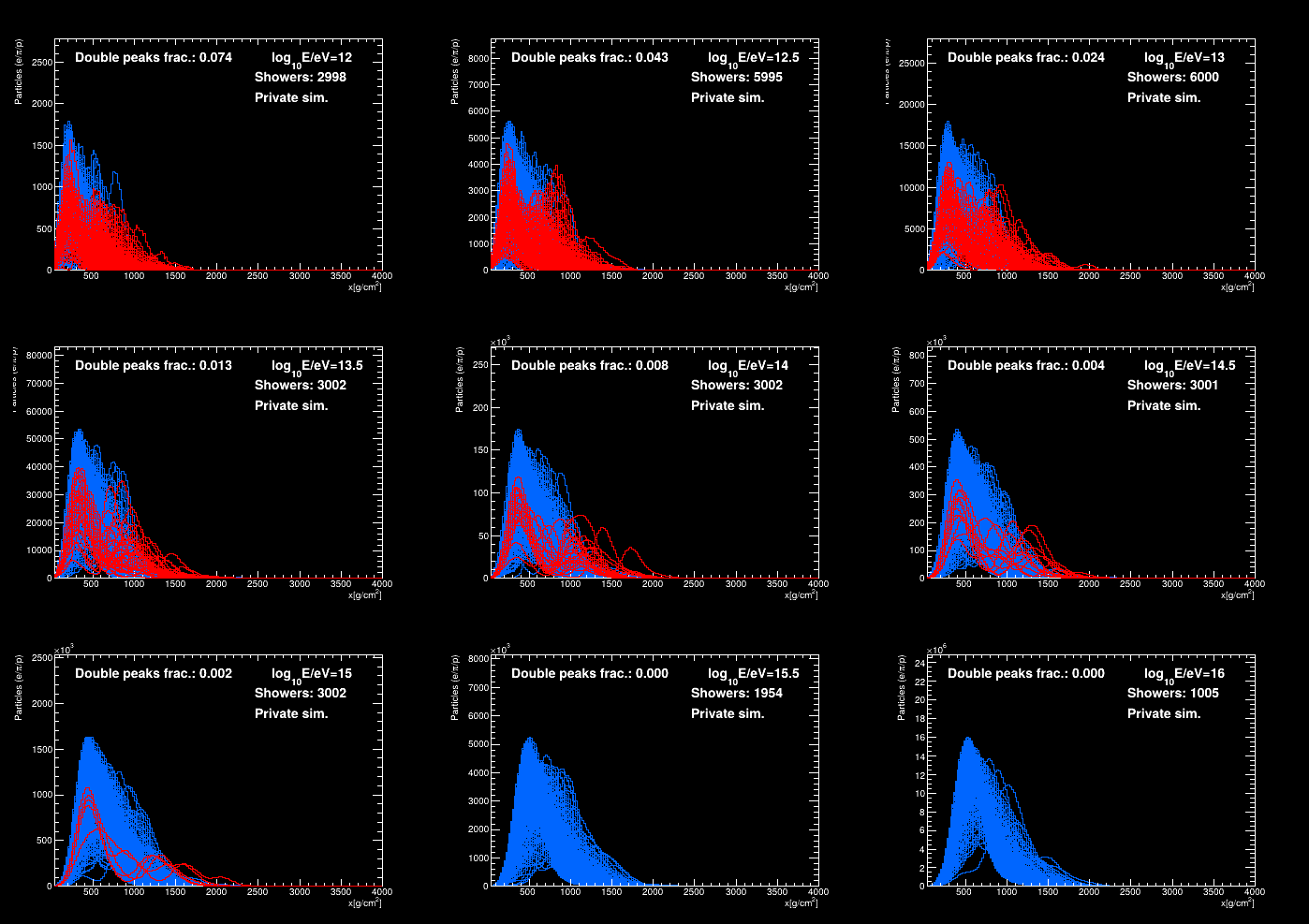} \\ 
    \caption{Example of 1k longitudinal shower profiles of the presented model in terms of the number of charged particles as function of the atmospheric depth. Showers of a larger standard deviation are shown in red, being suspects of double-peak structured showers.}
    \label{fig:private_shower_profiles}
\end{figure}

\begin{figure}[!h]
    \centering
    \includegraphics[width=1.0\linewidth]{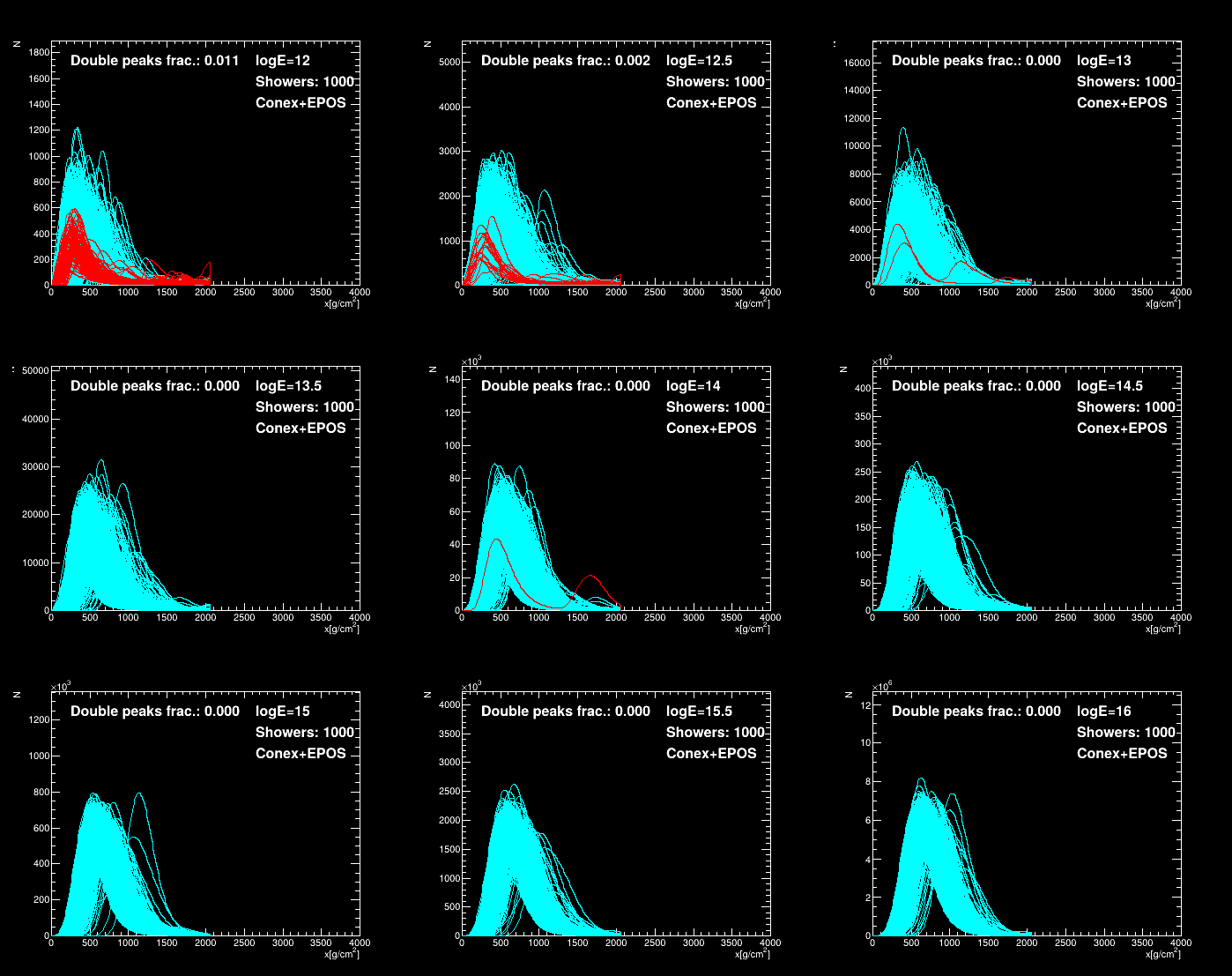} 
    \caption{Example of 1k longitudinal shower profiles of the \Conex{} generator with EPOS as the hadronization model left. Showers of a larger standard deviation are shown in red, being suspects of double-peak structured showers.}
    \label{fig:conex_profiles_epos}
\end{figure}

The multiple maxima appear with frequency of the order of percent or sub-percent level, see Figure~\ref{fig:private_shower_profiles} for the private simulation example, with the showers of non-standard profiles highlighted. Also the \Conex{} shower generator exhibits similar features as can be seen in~Figure~\ref{fig:conex_profiles_epos}.

\end{document}